\documentclass[aps,prx,reprint,
superscriptaddress,floatfix,longbibliography,footinbib]{revtex4-1}
\usepackage{mathtools}
\usepackage{amsfonts}
\usepackage{amsmath}
\usepackage{physics}
\usepackage{float}
\usepackage{sidecap}
\usepackage{bbold}
\usepackage[version=4]{mhchem}
\usepackage[doipre={DOI:~}]{uri}
\usepackage{dcolumn}% Align table columns on decimal point
\newcommand{\nocontentsline}[3]{}

\usepackage{xcolor}
\newcommand\hl[1]{\textcolor{black}{#1}}

\newcommand{\be}{\begin{equation}}
\newcommand{\ee}{\end{equation}}
\newcommand{\bea}{\begin{eqnarray}}
\newcommand{\eea}{\end{eqnarray}}
\newcommand{\ben}{\begin{equation*}}
\newcommand{\een}{\end{equation*}}
\newcommand{\ba}{\begin{align}}
\newcommand{\ea}{\end{align}}

\newcommand{\mbf}{\mathbf}
\newcommand{\mrm}{\mathrm}

\begin{document}

\title{An optical lattice with sound\\\small{(Nature \textbf{599}, 211 (2021); DOI:10.1038/s41586-021-03945-x)}}

\author{Yudan Guo}
\affiliation{Department of Physics, Stanford University, Stanford CA 94305, USA}
\affiliation{E.~L.~Ginzton Laboratory, Stanford University, Stanford, CA 94305, USA}
\author{Ronen M.~Kroeze}
\affiliation{Department of Physics, Stanford University, Stanford CA 94305, USA}
\affiliation{E.~L.~Ginzton Laboratory, Stanford University, Stanford, CA 94305, USA}
\author{Brendan P.~Marsh}
\affiliation{E.~L.~Ginzton Laboratory, Stanford University, Stanford, CA 94305, USA}
\affiliation{Department of Applied Physics, Stanford University, Stanford CA 94305, USA}
\author{Sarang Gopalakrishnan}
\affiliation{Department of Physics, The Pennsylvania State University, University Park, PA 16802, USA}
\author{\\Jonathan Keeling} 
\affiliation{SUPA, School of Physics and Astronomy, University of St. Andrews, St. Andrews KY16 9SS, United Kingdom}
\author{Benjamin L. Lev}
\affiliation{Department of Physics, Stanford University, Stanford CA 94305, USA}
\affiliation{E.~L.~Ginzton Laboratory, Stanford University, Stanford, CA 94305, USA}
\affiliation{Department of Applied Physics, Stanford University, Stanford CA 94305, USA}

\date{\today}

\begin{abstract}

Quantised sound waves---phonons---govern the elastic response of crystalline materials, and also play an integral part in determining their thermodynamic properties and electrical response (e.g., by binding electrons into superconducting Cooper pairs)~\cite{kittel2004its,Chaikin1995poc,tinkham2004its}. The physics of lattice phonons and elasticity is absent in simulators of quantum solids constructed of neutral atoms in periodic light potentials: unlike real solids, traditional optical lattices are silent because they are infinitely stiff~\cite{Grimm2000odt}. Optical-lattice realisations of crystals therefore lack some of the central dynamical degrees of freedom that determine the low-temperature properties of real materials. Here, we create an optical lattice with phonon modes using a Bose-Einstein condensate (BEC) coupled to a confocal optical resonator.  Playing the role of an \textit{active} quantum gas microscope, the multimode cavity QED system both images the phonons and induces the crystallisation that supports phonons via short-range, photon-mediated atom-atom interactions.  Dynamical susceptibility measurements reveal the phonon dispersion relation, showing that these collective excitations exhibit a sound speed dependent on the BEC-photon coupling strength. Our results pave the way for exploring the rich physics of elasticity in \emph{quantum} solids, ranging from quantum melting transitions~\cite{Beekman2017dgf} to exotic ``fractonic'' topological defects~\cite{Pretko2018fd} in the quantum regime.

\end{abstract}

\maketitle

\begin{figure*}[t]
\centering
\includegraphics[width = \textwidth]{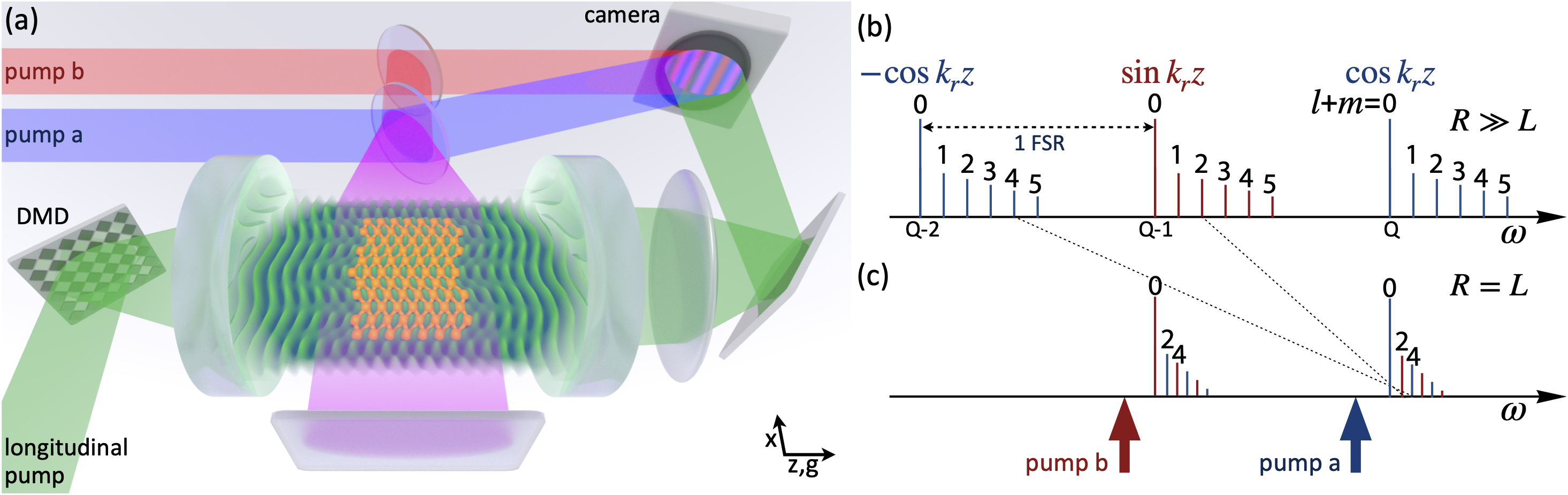}
\caption{\textbf{Transverse, double-pumped confocal QED system coupled to a BEC.} (a) Sketch of the vibrating atomic density wave (orange) created inside the confocal cavity field (green). Blue and red transverse pump fields are combined (purple) and retroreflected to form a standing wave. A beamsplitter directs some pump light onto a CCD camera to serve as a local oscillator (LO) for the holographic imaging of the spatially dependent phase and amplitude of the cavity emission. A digital micromirror device (DMD) injects patterns of light for measuring dispersion relations:  This light longitudinally pumps the cavity along $\hat{z}$ at frequency $\omega$ and transverse wavevector $k_\perp$ along $\hat{x}$. The momentum distribution of the atoms are measured in time-of-flight via absorption imaging along $\hat{x}$ (not shown). (b) Spectrum of a cavity whose radius of curvature $R$ is much greater than its length $L$, which thus operates in the resolved-mode regime of single or few-mode cavities.  Families of modes with fixed longitudinal mode number $Q$ have the same longitudinal phase offset; they differ in phase by $\pm\pi/2$ from families with $Q\pm1$ one FSR away. (c) The confocal cavity mode spectrum showing even parity families. (Frequency-degenerate transverse modes, labelled by the sum of their indices $l+m$, are dispersed in $\omega$ for ease of viewing.)  The modes in each family alternate in longitudinal phase. The blue and red arrows indicate the transverse pumps a and b, respectively.  }
 \label{fig1}
\end{figure*}

Ultracold neutral atoms confined in optical lattices have been a fruitful platform for ``emulating'' the itinerant motion of electrons in crystals~\cite{Bloch2008mpw}. Optical lattices, however, lack a crucial feature of real crystal lattices, which is \emph{elasticity}. Real crystal lattices vibrate, deform in response to electrons, and transmit stresses; in contrast, optical lattices are nondynamical. Elasticity has recently seen a revival of interest, motivated by developments such as ``fracton-elasticity duality''~\cite{Pretko2018fd}. While the motion of ions in real crystals is classical, a natural question is how elasticity would change in the presence of strong quantum zero-point motion of the atoms in the crystal; few controlled experimental studies of this regime exist. We realise a system that combines crystalline elasticity with quantum-degenerate motion, in the form of a compliant optical lattice arising from the crystallisation of Bose-condensed Rb atoms. While mimicking the effect of phonons in static optical lattices has been proposed~\cite{Gonzalez-Cuadra2018scb}, our method yields a continuum of phonon modes akin to those in solid-state materials.  

Crystallisation is the spontaneous breaking of the continuous translational symmetry of space. Due to this symmetry-breaking, a crystal has a manifold of physically distinct equilibrium states, which are related to one another by global translations (i.e., by sliding the entire crystal). Global, zero-momentum ($k = 0$) translations connect these equilibrium states and cost no free energy. Additionally, crystals with \emph{local} interactions have a continuum of finite-$k$ modes with arbitrarily low energies: these modes, called phonons, involve \emph{locally} sliding the crystal by an amount that varies slowly in space with a period $2\pi/k$.  Because global translations cost no energy, while local translations do, a crystal is \emph{rigid}, and responds globally to local stresses.  The phonon excitation branch, which is the Goldstone mode of the broken translational symmetry, governs the elastic properties of crystals. The properties of the phonon branch are intimately tied to those of topological defects, such as dislocations, which have recently been identified as ``fractonic'' excitations.  In contrast, for symmetry breaking arising from all-to-all interactions (or other sufficiently long-range interactions), while the $k=0$ zero mode may remain in place, there is a gap to all $k \neq 0$ excitations.  This gives topological defects an \emph{extensive} energy cost, and thus, in these long-range crystals, any nontrivial elastic response is frozen out.  (When the atoms forming a crystal are already Bose-condensed, so that the crystal is a ``supersolid,'' there are additional superfluid Goldstone modes. These superfluid modes are associated with the $U(1)$ phase of the condensate itself and also exist in the absence of a lattice~\cite{Stamper-Kurn1999eop,Pethick2002}.  Sound propagation and diffusion has been studied with strongly interacting fermions~\cite{Patel2020usd,Brown2018bmt}.)

The primarily contact interactions among Rb do not support crystallisation.
Optical cavity photons, however, can mediate interactions that do support crystallisation, as follows. We begin by considering a pump field oriented transverse to a Fabry-P\'{e}rot cavity axis that is far detuned from all but a single resonance. Above a critical threshold pump strength, a density wave (DW) polariton condensate~\cite{Carusotto2013qfo} forms via a superradiant (Hepp-Lieb-Dicke) phase transition:  The $N$ intracavity atoms cooperatively scatter pump photons into the cavity, forming a coherent optical state, while concomitantly the atoms adopt one of two chequerboard configurations of the $\lambda/2$-period lattice~\cite{Kirton2018itt,Mivehvar2021cqw}.  This $\cos{k_r x}\cos{k_r z}$ lattice is formed by the interference of the pump field with the emergent cavity field.  The two-photon scattering process excites the $k=0$ BEC into a superposition of $|k_x,k_z\rangle = |{\pm} k_r, {\pm} k_r\rangle$ momentum modes, where $\hat{x}$ ($\hat{z}$) is the pump (cavity) axis, the pump and cavity fields are of wavelength $\lambda \approx 780$~nm, and $k_r=2\pi/\lambda\approx8$~rad$/\mu$m is the recoil momentum; $\hbar\omega_r = \hbar^2k_r^2/2m\approx2\pi\hbar{\times}3.8$~kHz is the recoil energy, where $2\pi\hbar$ is  Planck's constant.
The state is stable if the pump frequency $\omega_P$ is red-detuned by $\Delta_C = \omega_P - \omega_C < 0$ from the cavity resonance $\omega_C$.  

Such single-mode experiments have enabled the exploration of a variety of quantum collective phenomena~\cite{Klinder2015ooa,Landig2016qpf,Kollar2017scw,Kroeze2019dsc}, but do not have a continuous translational symmetry and are thus neither rigid nor elastic. A continuous translational symmetry can be restored if one adds a second cavity mode.  \hl{For example, one emerges from the phase difference of forward- and backward-travelling waves in a ring-cavity geometry~\cite{Gopalakrishnan2009eca,Gopalakrishnan2010aco,Mivehvar2018dsi,Schuster2020spo}, or from the amplitude ratio when} crossing two cavities~\cite{Leonard2017sfi,Leonard2017mam}. \hl{In both these cases, this} gives rise to a continuous $U(1)$ family of steady states, which can be related by continuous global displacement of the atoms. We will explain this mechanism below. However, since in these experiments the interactions are mediated by only a few cavity modes, the interactions are infinite-range, so the ``crystal'' does not allow for nontrivial elastic deformations.

\begin{figure*}[t!]
\centering
\includegraphics[width = \textwidth]{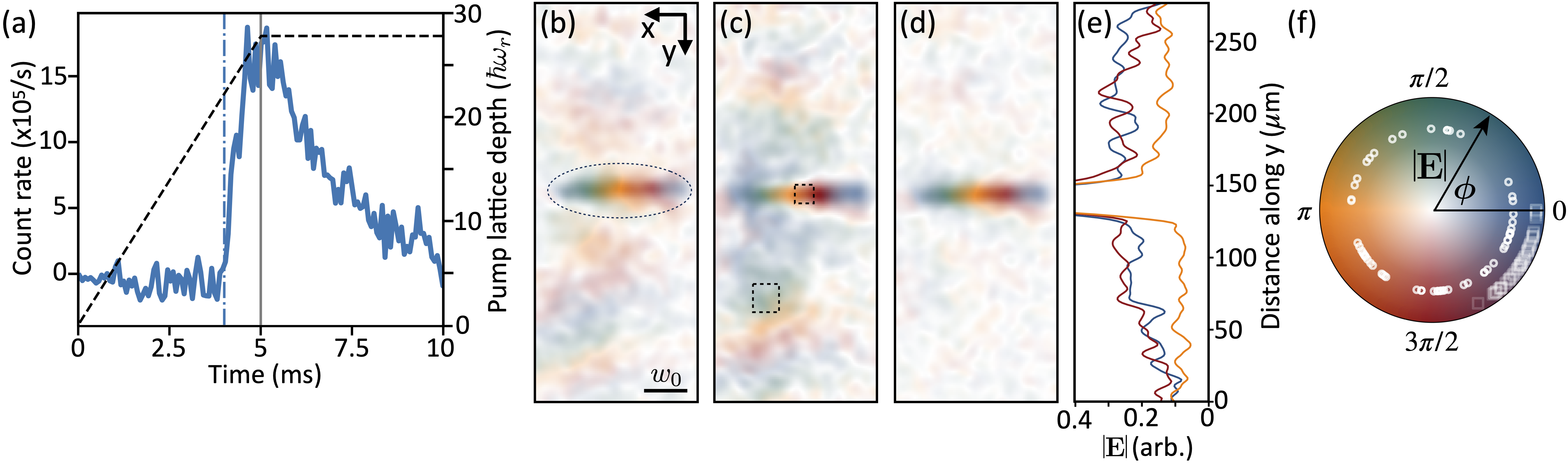}
\caption{\textbf{\hl{Efficacy of double-pumping scheme.}}   \hl{(a) Cavity emission intensity versus time (left axis); background counts have been subtracted.   The pump lattice depth ${\propto}\eta^2$ (right axis, dashed black line)  is linearly ramped through the superradiant threshold ($\eta^2=\eta_\text{th}^2$ at the blue vertical dash-dotted line) before being held constant at the measurement time (grey vertical solid line).}  (b-e) Evidence that the nonlocal interaction is cancelled by double-pumping.  Holograms of emission from a double-pumped confocal cavity with either (b) the blue pump LO or (c) the red pump LO illuminating the camera. The BEC image appears as the bright rainbow-like stripe. Its position in panel (b) is indicated by the dotted oval.  While this emission arises from the local interaction---the first term in Eq.~\eqref{eqn:intStrength}---emission outside this region is due to photons mediating the nonlocal interaction (the second term).  \hl{Note that the linear phase gradient (rainbow-like feature) in the local emission is an artefact caused by nonlinearities arising at strong pumping strengths:  While large strengths of  $\eta^2/\eta^2_\text{th}\approx10$ are needed to obtain high signal-to-noise in these images,  far weaker strengths $\eta^2/\eta^2_\text{th} \leq 1.25$, at which these nonlinearities are negligible, are sufficient for taking the dispersion data in Figs.~\ref{fig3} and~\ref{fig4}. See Supplementary Information for details.}  (d)  The absence of the nonlocal interaction is revealed by taking the digital sum of these holograms.   \hl{(e) This is more clearly shown by integrating each of these images along $\hat{x}$ and comparing the intensity level of nonlocal emission in the single-pumped traces [blue for panel (b), red for (c)] to the double-pumped trace (orange).} The emission from the nonlocal interaction is \hl{much reduced} due to the cancellation of the nonlocal parts in $U^\text{mm}_\text{a}$ and $U^\text{mm}_\text{b}$ when both pumps are present. \hl{ (f)  A scatter plot of the shot-to-shot phase of the DW polariton under the single (square) and double (circle) pumping scheme is overlaid on the colour scale for the field amplitude $|E|$ and phase $\phi$. This phase is the difference between the local and nonlocal phase in the emission regions indicated by, e.g., the dashed boxes panel (c).   (Symbols in each set are offset in radius for clarity; 60 points are shown for each.)  The near-random distribution about $2\pi$ for the double-pumping scheme, as opposed to the clumping of the single-pumped phases, illustrates the emergence of the $U(1)$ symmetry.  This demonstrates that in panels (d,e) the nonlocal interaction is sufficiently cancelled; see Supplemental Information for more details.  }}
 \label{fig2}
\end{figure*}

Coupling atoms to far more than two modes is necessary to create a compliant lattice that may, e.g., lead to superfluids with quantum liquid crystalline structure or exhibiting Meissner-like effects and Peierls instabilities~\cite{Gopalakrishnan2009eca,Gopalakrishnan2010aco,Ballantine2017mef,Rylands2020ppt}. One can superpose many degenerate modes to form compact supermodes---localised photon wavepackets~\cite{Kollar2017scw,Vaidya2017tpa}.  Exchanging these localised photons leads to finite-range interactions and momentum exchange~\cite{Vaidya2017tpa,Guo2019spa,Guo2019eab}. As such, when a multimode cavity is combined with a double-pumping scheme to engineer a $U(1)$ symmetry, a fully fledged Goldstone mode with a dispersion relation should emerge. If using a BEC, the result would be a supersolid with phonons.  (Phonons have also been proposed \hl{and sought via the refractive index change of atoms coupled to strong light fields ~\cite{Lewenstein2006tte,Ostermann2016sco,Dimitrova2017oot}} and exist as natural modes in ion traps~\cite{Monroe2021pqs}.)

In what follows, we first review how a $U(1)$ symmetry can be engineered by double-pumping in the context of a two-mode cavity~\cite{Guo2019eab}; we then extend it to multimode cavities. We consider two pump fields, labelled `a' and `b', each detuned by $\Delta_C$ from one of two cavity modes spaced one free spectral range (FSR) apart.  Figure~\ref{fig1}a sketches a transversely double-pumped cavity, while Fig.~\ref{fig1}b shows the spectrum of a cavity whose length $L$ greatly exceeds its mirrors' radius of curvature $R$. Each pump induces an interaction between atom pairs. The combined interaction takes the form:
\be
U^\text{sm}_\text{total} \propto U^\text{sm}_\text{a}\cos{k_rz}\cos{k_rz^\prime} + U^\text{sm}_\text{b}\sin{k_rz}\sin{k_rz^\prime},
\ee
where $U^\text{sm}_i$\hl{$(x,x')$} $ = \eta_i^2\cos{k_rx}\cos{k_rx^\prime}/\Delta_C$ is the interaction strength induced by each pump field of intensity $\propto\Omega_i^2$ \hl{and is negative under red detuning}. \hl{The two-photon coupling is $\eta_{i} \equiv g_0 \Omega_i/4\Delta^{i}_A$, and for notational simplicity, we will drop the $(x,x')$ arguments of $U^\text{sm}_i$.} The coupling strength of a single atom to the cavity field is $g_0$, and the pumps are detuned from the atomic level by $\Delta^{i}_A$, which are much greater than $\Delta_C$ and the transition linewidth.  The change from cosine to sine reflects the shift by  $\lambda/2$ in the longitudinal field profile as the mode's longitudinal index $Q$ changes by one; $Q$ is the number of optical half-wavelengths separating the mirrors.  
The interaction strengths become equal when \hl{$\eta_\text{a} = \eta_\text{b} \equiv\eta$}, with the result that $U^\text{sm}_\text{a} =  U^\text{sm}_\text{b}\equiv U^\text{sm}$ and the interaction has a continuous translational symmetry along $\hat{z}$: $U^\text{sm}_\text{total} = U^\text{sm}\cos[k_r(z-z^\prime)]$. 

\begin{figure*}[t!]
\centering
\includegraphics[width = \textwidth]{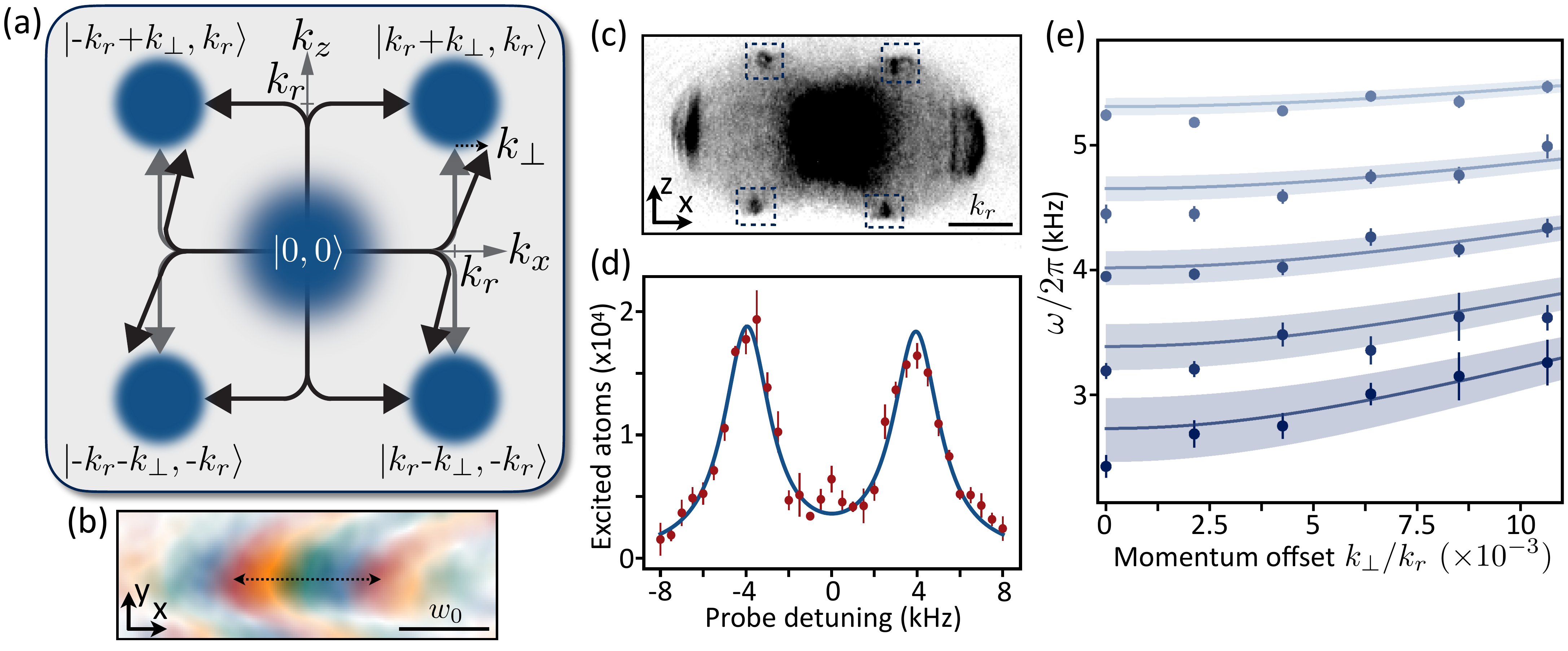}
 \caption{\textbf{Soft-mode dispersion of density-wave polaritons below threshold.} (a) The two-photon scattering process excites atoms by receiving one momentum  kick from a pump photon along $\pm\hat{x}$ and another from a cavity photon along $\pm\hat{z}$.  The transverse momentum of the higher-order cavity modes shifts the $\pm\hat{x}$ momentum by an amount $\pm k_\perp$. Shown is one possible momentum state; Supplementary Information describes the others.  (b) While transversely pumping below threshold, we stimulate a soft mode with a particular $k_\perp$ by seeding the cavity longitudinally.  An example of a seed field is shown here, as imaged by the transmission from an empty cavity. Cavity and imaging distortions curve the  $k_\perp x$ phase fronts.    (c) Below-threshold absorption image of atoms in time-of-flight after a small fraction have been Bragg-stimulated into the four peaks indicated by squares. The other two peaks arise from the pump lattice alone. (d) Example Bragg scattering spectrum showing the number of scattered atoms versus probe detuning from \hl{$\omega_P$}. Data are the sum of atoms within the squares in panel (c) for $\eta^2/\eta^2_\text{th} =0.5$ and $k_\perp/k_r \approx2.5\times10^{-3}$.  A double Lorentzian (blue curve) is fit to the data, and the excitation energy is half the separation between peaks. Vertical bars represent standard error here and below. (e) Dispersion relation \hl{$\omega$ versus $k_\perp$ for momenta offset from zero by $k_r$.   The curves are plotted for} pump strengths ranging from \hl{$\eta^2/\eta^2_\text{th}=0.3$ at the top (lightest blue) to 0.7 at bottom (darkest blue)} in steps of 0.1.  Each column of data at fixed $k_\perp$ shows a softening roton mode as the supermode DW polariton condenses at threshold. Data are compared to parameter-free theory curves (with error bands) derived from the theory developed in the Supplementary Information.}
\label{fig3}
\end{figure*}

To create a lattice where phonons exist, we extend this double-pumping scheme to a multimode cavity of confocal configuration ${L=R=1}$~cm. Figure~\ref{fig1}c shows the confocal mode spectrum, which contains families of degenerate modes of either all-even or all-odd parity~\cite{Siegman1986l}.  We pump two even mode families spaced one FSR apart.  As in the single-mode case, the longitudinal mode profile alternates between cosine and sine.  In addition, higher-order transverse modes in the \textit{same} family also alternate in longitudinal profile. The interactions driven by the pumps is therefore $U^\text{mm}_\text{total} \propto U^\text{mm}_\text{a} + U^\text{mm}_\text{b}$, where
\bea
U^\text{mm}_\text{a} = U_0\cos{k_rz}\cos{k_rz^\prime} +  U_2\sin{k_rz}\sin{k_rz^\prime}\label{eqn:IntA} \\
U^\text{mm}_\text{b} = U_0\sin{k_rz}\sin{k_rz^\prime} +  U_2\cos{k_rz}\cos{k_rz^\prime}\label{eqn:IntB}.
\eea
As discussed in the Supplementary Information, the interaction strengths $U_{\{0,2\}}$ are~\cite{Vaidya2017tpa,Guo2019spa,Guo2019eab}
\be\label{eqn:intStrength}
U_{\{0,2\}}/U^\text{sm} 
%&=& \mkern-30mu\sum_{l,m:l+m= \{0,2\}\, \text{mod}\, 4}\mkern-40mu \frac{\Xi_{l,m}(\mathbf{r})\Xi_{l,m}(\mathbf{r}^\prime)}{\hl{1-(l+m)\epsilon/\Delta_C}} \nonumber\\
\approx \frac{e^{-\Delta r/\xi}}{\sqrt{\Delta r/\xi}} \pm \cos\left[\frac{\mathbf{r}\cdot{\mathbf{r}^\prime}}{w_0^2/2}\right].
\ee
%Here, $l,m$ index  the transverse electromagnetic modes TEM$_{l,m}$ with Hermite-Gaussian mode functions $\Xi_{l,m}(\mathbf{r})$. 
The waist \hl{of the lowest-order cavity mode is} $w_0 {=} 35$~$\mu$m.  The first term is the local interaction with $\Delta r {=} |\mbf{r}-\mbf{r}^\prime|$ and a range $\xi\agt 2$~$\mu$m; \hl{$\xi \approx 5$~$\mu$m for the data presented and} is set by the number of degenerate modes supported by the confocal cavity and $\Delta_C$~\cite{Vaidya2017tpa,Kroeze2021preprint}.  The second term is the nonlocal interaction.  It cancels in the double-pump scheme, yielding both the desired local interaction and the $U(1)$ translational symmetry:
\be\label{UmmTotal}
U^\text{mm}_\text{total} = U^\text{sm}\frac{e^{-\Delta r/\xi}}{\sqrt{\Delta r/\xi}}\cos{k_r\Delta z}.
\ee
\hl{A local mirror image term, omitted above, does not play a role in this work because we place the atoms in only one half-plane of the cavity.}

We demonstrate the cancellation of the nonlocal contribution to the cavity-mediated interaction by imaging the phase and amplitude of the above-threshold superradiant emission under double-pumping; see Supplementary Information for details. Figures~\ref{fig2}b,c show the cavity emission from photons mediating $U^\text{mm}_\text{a}$ and $U^\text{mm}_\text{b}$, respectively.  The local interaction created by the atoms gives rise to the image of the BEC in the emergent lattice.  That is, the interaction manifests as an emitted image because it is the local light in the cavity that mediates the interaction and this same light leaks out of the cavity~\cite{Vaidya2017tpa}; see Supplementary Information.  The emission surrounding the BEC is from the nonlocal interaction~\cite{Guo2019spa,Guo2019eab}. This nonlocal interaction is cancelled under double pumping. This  manifests as an image without nonlocal emission $\propto U^\text{mm}_\text{total}$, which we can obtain through the digital summation of the single-LO images.  Indeed, this is what we observe in Fig.~\ref{fig2}d \hl{and in the line integrations of Fig.~\ref{fig2}e. The resulting emergence of the $U(1)$ symmetry manifests as a random distribution of DW phases each time the system is pumped above threshold.  This is shown in Fig.~\ref{fig2}f; see Supplementary Information for measurement procedure and discussion. A representative time trace of cavity emission is shown in Fig.~\ref{fig2}a.}

\hl{As a first study of this translationally symmetric system, we focus on the below-threshold spectroscopy of the $k$-dependent, normal (roton) dispersion where there is no lattice and thus no phonon.  We will then show how the excitation spectrum changes above threshold in the presence of the emergent lattice. }  In single-mode cavities, a roton instability at $k = k_r$ results in a DW polariton condensate at threshold~\cite{Mottl2012rms}.  In a confocal cavity, by contrast, \hl{momentum-exchange mediated by} the local interaction allows atoms to scatter into a range of states with $|k_\perp|$ added to $|k_r|$ along $\pm\hat{x}$, as illustrated in Fig.~\ref{fig3}a. Consequently, the \textit{supermode} DW polariton shows broad roton minima \hl{softening} near $k_r$.  We can probe the dispersion around this point by stimulating the cavity at a particular $k_\perp \geq 0$.  This is done by injecting a longitudinal pump field whose amplitude and phase has been programmed by the DMD to be ${\propto} e^{ik_\perp x}$~\cite{Papageorge2016ctm}; see Fig.~\ref{fig1}a for illustration.  Figure~\ref{fig3}b shows an example field pattern. We can stimulate values close to the characteristic momentum scale of this multimode cavity $\zeta\equiv \xi^{-1}\approx 0.02 k_r$; see Supplementary Information.

We measure the dispersion of these $k$-dependent roton soft modes by cavity-enhanced Bragg stimulation. While pumping below threshold, we stimulate the cavity with the longitudinal probe field at a particular $k_\perp$ set by the DMD. Atoms are more efficiently scattered into the Bragg peaks of Fig.~\ref{fig3}a when the frequency and wavevector of the probe field match the roton dispersion, which varies with pump strength \hl{$\propto\eta$}.   We directly absorption-image these scattered atoms in time-of-flight, as shown in Fig.~\ref{fig3}c. Summing the atoms in all four peaks, we can plot the excitation spectrum for a given $k_\perp$ and \hl{$\eta$}; see Fig.~\ref{fig3}d.  Figure~\ref{fig3}e compiles the excitation frequencies.  As \hl{$\eta$} increases, we see the rotons soften and become more strongly dispersive---i.e., display a stronger $k$ dependence.  At \hl{$\eta$} $\to 0$, the dispersion is that of atomic DWs set by the atomic mass; with increasing \hl{$\eta$}, atomic DWs mix with photons to form DW-polaritons leading to a steeper dispersion; see Supplementary Information.  

\begin{figure}[t!]
\centering
\includegraphics[width = \columnwidth]{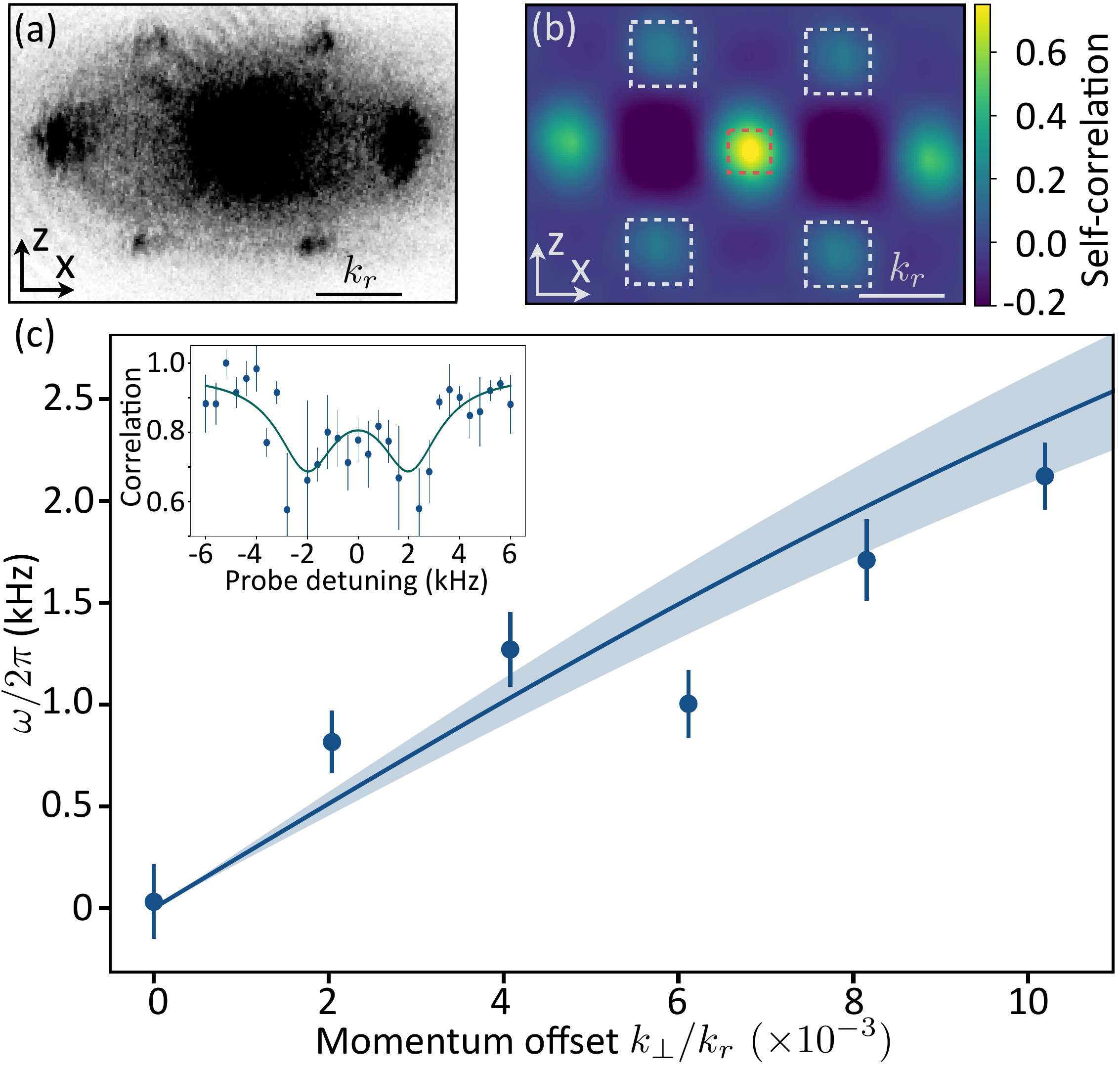}
\caption{\textbf{Goldstone dispersion relation $\omega(k_\perp)$.} 
(a) Example above-threshold time-of-flight image recording the momentum distribution $\rho(k)$. (b) The self-correlation analysis yielding $\langle \rho(k+\delta k)\rho(k)  \rangle$. The white dashed squares indicate regions of interest for extracting the correlation strength associated with the chequerboard lattice. The correlation strength is calculated by normalising the sum of the values in white squares by that in the central red dashed square. (c) The dispersion relation curve (blue) is overlaid using the theory presented in the Supplemental Information and is parameter-free. The data are \hl{ plotted with momenta offset from zero by $k_r$. They are} consistent with a linear dispersion for $k_\perp {\ll} \zeta$. The pump strength is \hl{$\eta^2/\eta^2_\text{th} =1.25$}.  The error band (light blue) represents one-sigma error in the theory parameters.  Inset shows an example dynamic susceptibility spectrum obtained from  self-correlation analyses  taken for $k_\perp/k_r \approx 0.01$.  Correlations decrease on resonance because the momentum structure factor differs from the $|0,0\rangle$ state due to the addition of $k_\perp$.}
\label{fig4}
\end{figure}

\hl{We now demonstrate that the dispersion of the lattice phonon branch is gapless and linear. We do so by measuring} the Goldstone dispersion of the phonon modes of the lattice that forms above threshold. We again use cavity-enhanced Bragg stimulation to measure dynamic susceptibility $\omega(k_\perp)$.  Above threshold, however, the Bragg peaks of the DW polariton condensate are too populous to discern the additional Bragg-scattered atoms.  Moreover, $k_\perp \ll k_r$ and our maximal time-of-flight is too short to discern the additional $\pm k_\perp$ from the spread in $k$ about the peaks.  We instead employ a self-correlation analysis of the momentum distribution to extract the phonon mode resonances; see Supplementary Information. For example, Fig.~\ref{fig4}b is the self-correlation of the momentum image in Fig.~\ref{fig4}a for a particular $\omega$ and $k_\perp$.  The result is the Goldstone mode dispersion curve in Fig.~\ref{fig4}c. As derived in the Supplementary Information, the low-$k$ dispersion is linear: $\omega(k_\perp) = v_s|k_\perp|$, with a sound speed  $v_s \hl{ \simeq \sqrt{\hbar \omega_r[1+(\eta/\eta_{\text{th}})^2](1/2m+  E_I/\zeta^2)}}$, where the cavity-mediated interaction strength is $E_I=-8\eta^2 N/\Delta_C$.  One may note that typically $E_I/\zeta^2 \gg 1/2m$, so the phonon dispersion is predominantly set by the cavity-mediated interactions. These phonon excitations have a sound velocity of 16~cm/s, $10^4\times$ slower than that in, e.g., copper at room temperature. The theory curve contains no free parameters. 

The self-consistent cavity optical lattice and atomic DW create a compliant lattice, while individual atoms remain itinerant within the lattice. \hl{In real space, the acoustic phonon modes manifest as a sliding motion of the lattice along the cavity axis $\hat{z}$, modulated  along the $\hat{x}$ direction with wavevector $k_\perp$, corresponding to a transverse phonon mode; this motion is illustrated in the Supplementary Video. (Here, we use ``transverse'' to refer to the relative directions of displacement versus wavevector, rather than with respect to the cavity axes.)}  

\hl{Note that instead of imaging the momentum distribution, we could also have imaged the phonon through the pattern of light emitted from the cavity (see Supplementary Information for an example).} In this sense our system acts as an unusual ``active" quantum gas microscope, in which the cavity fields mediate interactions that support phonons, while their emission provides spatial information about the atomic density profile. 

Adding other atomic spin states or species within this dynamic lattice would more directly mimic electrons in traditional solid-state systems. Replacing the BEC with a degenerate Fermi gas might provide opportunities to study the electron-phonon physics related to polarons~\cite{Devreese2009fpa} in a context complementary to previous studies~\cite{Hu2016bpi,Jorgensen2016ooa,Yan2020bpn} or to study metallic transport in strange metals beyond the semiclassical approximation of long-lived quasiparticles~\cite{Werman2017nta}. Moreover, the dual role of itinerant coherent atoms forming a compliant dynamical lattice may also provide access to regimes not attainable in solid-state systems, e.g., to resolve phonon number states in order to perform quantum acousto-optical experiments with supersolids.

We thank Steve Kivelson, Sean Hartnoll, and Vedika Khemani for stimulating discussions. We acknowledge funding support from the Army Research Office.  Y.G.~and B.M.~acknowledge funding from the Stanford Q-FARM Graduate Student Fellowship and the NSF Graduate Research Fellowship, respectively. S.G. acknowledges support from NSF Grant No. DMR-1653271. 

 %\bibliographystyle{apsrev4-1-prx}
 %\bibliography{cQED,LevLab}


%%%%%%%%%%%%%%%%%%%%%%%%%%%%%%%%%%%%%%%%%%%%%%%%%%%
\clearpage

\onecolumngrid

\begingroup
\let\addcontentsline\nocontentsline
\section*{Materials and Supplementary Information}
\endgroup

\tableofcontents

\setcounter{figure}{0}
\setcounter{equation}{0}  
\renewcommand{\thefigure}{S\arabic{figure}}
\renewcommand{\theequation}{S\arabic{equation}}

\section{BEC preparation}
Bose-Einstein condensate production proceeds as in Ref.~\cite{Kollar2015aac}. To shape the BEC for this experiment, we use the same dynamical trap shaping technique as employed in our previous work reported in Ref.~\cite{Guo2019eab}.  A nearly pure BEC is created in state $| F = 1, m_F = -1\rangle$.   A harmonic potential consisting of two crossed beams of wavelength 1064 nm forms a trap of frequencies $(\omega_x,\omega_y,\omega_z)=2\pi\times[52.6(2),52.8(2),91.5(4)]$~Hz. The BEC population is $N=4.1(3)\times10^5$ and has  Thomas-Fermi radii of $(R_x,R_y,R_z)=[12.3(2),12.2(2),7.1(1)]$~$\mu$m. Finally, by changing the dither pattern of the trapping beams perpendicular to the pump, the trap shape is adiabatically deformed to produce an elongated gas of 93~$\mu$m along the pump direction $\hat{x}$. A harmonic potential in the other two directions is maintained with similar trap frequencies in the other two directions. The centre-of-mass of its density distribution lies at $\mbf{r}_{\mathrm{cm}} = (\text{49~}\mu\mathrm{m},\text{35~}\mu\mathrm{m})$ along $\hat{x}$ and $\hat{y}$ with respect to the cavity centre. 

\section{Cavity, pump lasers, and frequency locks}

The confocal cavity is vibrationally stabilised using the method presented in Ref.~\cite{Kollar2015aac}.  It is 1-cm long and has a radius of curvature $R =1$~cm, resulting in a  waist of its TEM$_{0,0}$ mode of $w_0=35$~$\mu$m.  Its finesse is $5.5 \times 10^4$, yielding a cavity linewidth of $\kappa = $137~kHz.  With a single-atom, single-mode coupling $g_0$ of $2 \pi \times1.47$~MHz, the single-atom, single-mode cooperativity is $C = 2g_0^2/\kappa\gamma = 5$, where the atomic linewidth is $\gamma = 2\pi\times6$~MHz.  Assuming a supermode enhancement factor of $\sim$10 (proportional to the inverse local interaction length scale $\xi$)~\cite{Vaidya2017tpa,Kroeze2021preprint}, the supermode single-atom cooperativity is $C^*\approx50$.  

\begin{SCfigure}
    \centering
    \includegraphics[width = 0.5\textwidth]{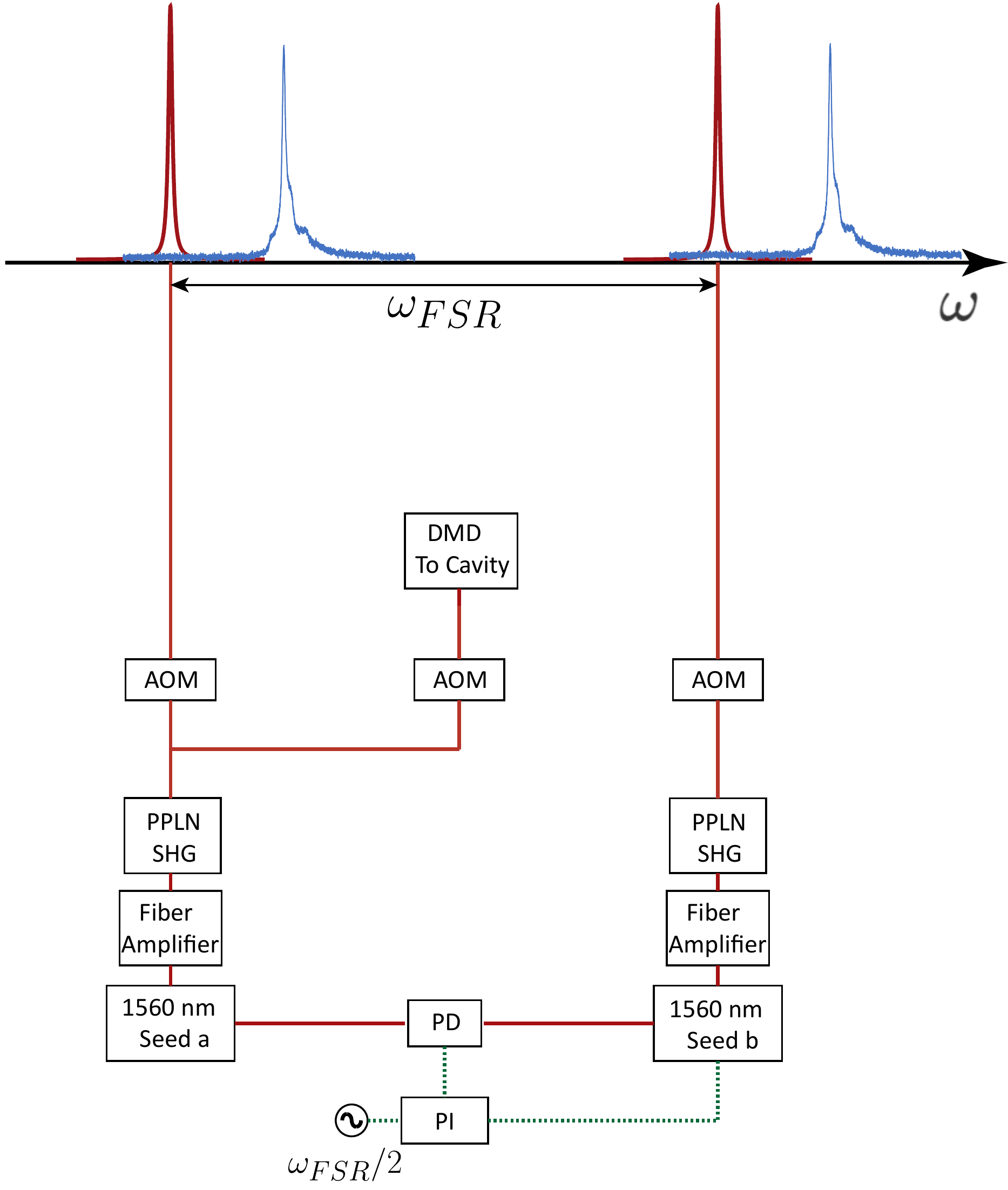}
    \caption{Pumping and laser locking schematics for two pump fields separated by one free spectral range (FSR).  Blue trace is the experimentally measured confocal transmission, while red curve is an illustration of the pump field line shape detuned by $\Delta_C$ from the nearby cavity resonance in blue.  Listed are a proportional-integral (PI) loop filter, photodetector (PD), periodically poled lithium niobate (PPLN) doubling crystals for second harmonic generation (SHG), acousto-optical modulators (AOMs), and a digital micromirror device (DMD).}
    \label{pumping}
\end{SCfigure}

The 780-nm pump beams are each derived from a frequency-doubled 1560-nm fiber amplifier and seed laser; see Fig.~\ref{pumping}. The relative frequency between the two 1560-nm seed lasers is stabilised with respect to a frequency source oscillating at half of the cavity free spectral range  ${\sim}7.5$~GHz. This frequency difference is controlled using a proportional-integral loop filter with feedback applied to seed `b'. A portion of the doubled 780-nm light from seed `a' is used as the illumination beam for the digital micro-mirror device. The DMD reflects this light into the path of the longitudinal cavity injection beam. Acousto-optical modulators are used to stabilise the intensity and adjust the relative detuning between the beams.   Additional 1560-nm light from seed `a' is used to stabilise the science cavity using the Pound-Drever-Hall technique. The two pumps are detuned from the 5$^{2}S_{1/2}|2,-2\rangle$ to 5$^{2}P_{3/2}$ transition by 96~GHz and 111~GHz, respectively. Throughout the experiments, the pumps are equally detuned from the relevant cavity resonances by $\Delta^\text{a}_C = \Delta^\text{b}_C \equiv \Delta_C = -2 \pi \times 50~\mathrm{MHz}$.

\section{Lattice calibration and pump balancing}
We calibrate the lattice depth of pump beams by performing Kapitza-Dirac diffraction of the BEC. The phase of the pump fields at the BEC is controlled by the retroreflection mirror shared by the pump beams. Measuring the lattice depth of the combined pump beams, we adjust the translation stage on which this mirror is mounted to match the phases of the pump lattices at the position of the atoms. We note that the beat length of the two pump lattices (separated in optical frequency by 15~GHz) is $\sim$5~mm, much larger than the atomic cloud size. Therefore, small mechanical fluctuations from the mirror mount will not cause the lattice to become out-of-phase at the atoms. The difference in recoil energy from this difference in frequency  is on the order of ${\sim} 0.1$~Hz and thus negligible, as is the change in wavelength. 

To bring into balance the cavity-mediated interactions induced by each pump, we perform a sequence of single-pump self-organization experiments.   We linearly ramp-up each beam in 5~ms and note the time at which the superradiant threshold is reached. The interaction strength can then be balanced by adjusting the ramp rate such that superradiance on a single FSR occurs at the same time for each beam. This ensures that the Raman coupling rate from each FSR is balanced, i.e, $\eta_a = \eta_b $, which then balances the cavity interaction strength for each pump.

\section{Holographic reconstruction of cavity emission}\label{holosec}
To perform the holographic imaging (spatial heterodyne detection) of the cavity emission, we follow the procedure established in Refs.~\cite{Kroeze2018sso,Guo2019spa} for a single pump field and extend it to the case of two pumps. Above threshold, the cavity emission has optical frequency content at both $\omega_a$ and $\omega_b$ (the two pump frequencies), separated by one FSR. To fully reconstruct the cavity electric field, therefore, one must illuminate the camera with two large local oscillator (LO) beams at frequencies $\omega_a$ and $\omega_b$ at different angles with respect to the propagation direction of the cavity emission.  This is illustrated in Fig.~\ref{fig1}a. The interference between LO and the cavity emission produces an image with an intensity $I_h(\mbf{r})$ that may be expressed as
%\begin{widetext}
\begin{align}
I_{h}(\mbf{r}) = \sum_{i = a,b}\lvert E_{c,i}(\mbf{r}) \rvert ^2 + \lvert E_{\text{LO},i}(\mbf{r}) \rvert ^2 + 2\chi_i |E_{c,i}(\mbf{r})E_{\text{LO},i}(\mbf{r})| \cos \left[ \Delta \mbf{k}_i \cdot \mbf{r} + \Delta\phi_i(\mathbf{r}) + \delta_i \right],
\label{hologram}
\end{align}
%\end{widetext}
where we have ignored the fast oscillating term at  $\omega_b - \omega_a$, and $E_{c,i}$ and $E_{\text{LO},i}$ are the cavity fields and LO fields for the two FSRs, respectively. Reduction of fringe contrast is characterised by the factor $\chi_i$. The additional phase terms $\delta_i$ accounts for the overall phase drift between the LO beams and the cavity emission in each experimental realisation due to technical fluctuations of the apparatus. Because of the angle difference, information from the cavity fields $E_{c,a}$ and $E_{c,b}$ are encoded in spatial wavevectors $\Delta\mbf{k}_a$ and $\Delta\mbf{k}_b$, respectively. Assuming the cavity field varies slowly over the spatial scale $2 \pi/|\Delta \mbf{k}_i|$, we may then extract the cavity field amplitudes $|E_{c,i} (\mbf{r})|$ and phase profiles $\Delta\phi_i(\mathbf{r}) + \delta_i$ by demodulating the image at $\Delta \mbf{k}_{i}$.

By using this scheme---an LO at each frequency but at different spatial wavevectors---we take a single spatial heterodyne image that simultaneously allows us to reconstruct the intracavity field for each resonance.   The phase of the nonlocal emission should differ by $\pi$ in the two images and indeed this signal cancels in their digital sum, as shown in Figs.~\ref{fig2}d,e.

\section{Generation of longitudinal probe with the DMD}

\begin{figure*}[b!]
  \includegraphics[width=0.99\textwidth]{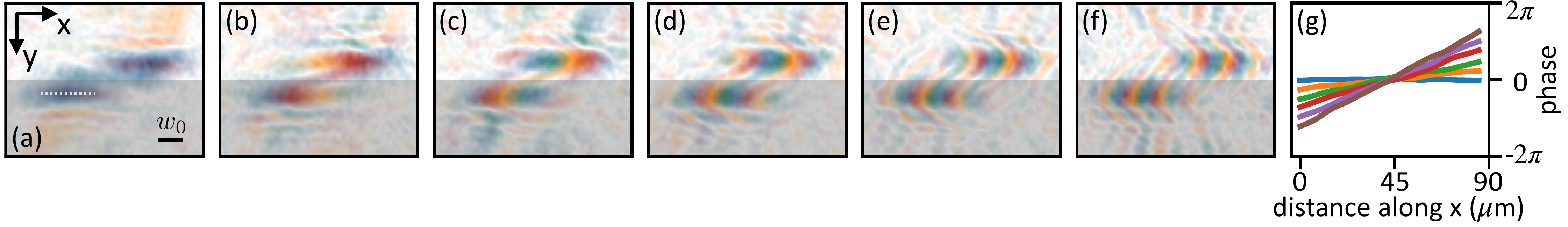}
  \caption{Measured DMD probe transmission cavity field and their phase profile line cuts. The values of $k_\perp/k_r$ in panels (a)--(f) are $[0, 2.1, 4.2, 6.3, 8.5, 10.6]\times 10^{-3}$, respectively. The white dashed line in panel a shows the length of the cuts in panel g. Additional features around the main probe field are due to imperfections of the confocal cavity and stray light from the DMD probe beam. The grey area is the half plane that contains the mirror image of the probe field, and we do not show this redundant portion of the image in the main text figures.}
  \label{SuppFig2}
\end{figure*}

The DMD plane is set at approximately the Fourier plane of the cavity centre by using a 100-mm focal length in-vacuum plano-convex lens. The phase aberration of the DMD and misalignment of the illumination beam must be calibrated out of the field images sent into the cavity.  We first calibrate these aberrations with an out-of-vacuum setup, similar to that used in Ref.~\cite{Papageorge2016ctm}. Then, using a cavity that is far from the confocal degeneracy point, an additional quadratic phase correction is added onto the DMD transfer function to effectively bring the DMD plane to the Fourier plane of the cavity centre. Finally, any intracavity field we desire can be generated by programming its Fourier transform to be displayed on the DMD. In our experiment, we perform Bragg spectroscopy at six different momenta; the measured DMD probe fields associated with these momenta are shown in Fig.~\ref{SuppFig2}.  The maximum $k_\perp$ modulation we can inject is limited by the numerical aperture of the lens that in-couples the DMD light and by the holding piece of the mirror.

\section{Bragg spectroscopy and self-correlation analysis}\label{dataanalysis}

The dynamic susceptibility of the system can be measured by using the longitudinal probe imprinted with a phase modulation $\propto k_\perp$ along $\hat{x}$ to stimulate, along with the pump fields, the scattering of atoms into the momentum states  $|\Psi(k_\perp)\rangle_+ = \sum_{\sigma,\tau=\pm 1} |\tau k_r + \sigma k_\perp, \sigma k_r\rangle$, as illustrated in Fig.~\ref{fig3}a. There is another possible set of states that we do not choose to stimulate or imprint given by  $|\Psi(k_\perp)\rangle_- = \sum_{\sigma,\tau=\pm 1} |\tau k_r - \sigma k_\perp, \sigma k_r\rangle$; note that  $|\Psi(-k_\perp)\rangle_+=|\Psi(k_\perp)\rangle_-$.  We choose $|\Psi\rangle_+$ versus $|\Psi\rangle_-$ by setting the phase of the field imprinted by the DMD.  The $|\Psi\rangle_+$ state yields the phase \textit{advancing} images in the main text. Because the scattering is coherent, the total atomic state is in a superposition of $|\Psi(k_\perp)\rangle_+$ and $|0,0\rangle$. In real space, adding the $|\Psi\rangle_+$ excitation on top of a uniform chequerboard lattice corresponds to adding a shearing lattice distortion.

We perform Bragg spectroscopy of the system's excitations by monitoring the increase in the population of the scattered atoms in the time-of-flight images versus the relative detuning between the longitudinal probe and the transverse pump. This detuning is adjusted with an AOM on the longitudinal probe beam path. The pump power is first ramped up to prepare the system with a given cavity-mediated interaction strength, and then the longitudinal probe beam is pulsed on for 0.5~ms. For measurements of mode-softening below the transition threshold, the response can be read-out by directly counting the atom population excited into the $|\Psi\rangle_+$ momentum state.  There are no background atoms at these momenta because there is no population of this momentum state in the normal phase: any atom signal is due to the Bragg excitation.  The resonance frequency is extracted by fitting the spectrum to a symmetric double-Lorentzian peak. The set of such frequencies is plotted in Fig.~\ref{fig3}e along with curves produced using the theory presented below.  The blue uncertainty bands are primarily due to the atom number uncertainty in the cavity-mediated interaction strength. The bands broaden close to threshold where the photon contribution plays an increased role.  

For measurements above the threshold, however, the situation is complicated by the macroscopic population of atoms already in the $|\Psi\rangle_0 = \sum_{\sigma,\tau=\pm 1} |\tau k_r, \sigma k_r\rangle$  excited momentum state. While the longitudinal probe creates an additional momentum excitation, the additional atoms are hard to distinguish from that already present because a) the number of these atoms is small compared to the number already condensed into this state, and b) $k_\perp \ll k_r$,  so that $|\Psi\rangle_+$ cannot be distinguished from $|\Psi\rangle_0$ given the limited 20~ms time-of-expansion of the time-of-flight image.  Thus, the same momentum-space atom-counting method used for below-threshold spectroscopy measurements is not viable.

We therefore turn to an alternative method that uses these same absorption time-of-flight (TOF) images, but performs an analysis based on momentum correlations rather than momentum-space atom counting.  To explain how this works, we first note that in real space, the longitudinal probe creates a small periodic distortion in the originally perfect chequerboard lattice.  We can quantify this distortion by computing the momentum-space self-correlation $\langle \rho(\mbf{k}+\delta \mbf{k})\rho(\mbf{k})  \rangle$ of the atomic momentum distribution $\rho(\mbf{k})$, which can be computed from $\langle \rho(\mbf{r}+\delta \mbf{r})\rho(\mbf{r})  \rangle_{\mathrm{TOF}}$ in each TOF image.  By focusing on the correlation between the shape of the wavepackets centred at momentum states $|\Psi\rangle_0$ and $|0,0\rangle$, we can discern the presence of atoms excited to $\pm k_\perp$ states.  This is because the correlation in the shape of $\rho(\mbf{k})$ at  $\mbf{k}=(0,0)$ and at the four $(\pm k_r,\pm k_r)$ regions is strongest when a perfect chequerboard lattice is present:  the wavepacket of the excited momentum state $|\Psi\rangle_0$ is simply a momentum displacement of that at $\mbf{k}=(0,0)$. However, in the presence of a small lattice distortion given by $k_\perp$, the structure factor is reduced and destructive matter-wave interference results in a reduction in the correlation. This correlation reduction is what is plotted in the inset of Fig.~\ref{fig4}.  The phonon mode resonances are manifest in the correlation signal dips.

To perform the above-threshold measurement, we first fit the entire image to a broad 2D Gaussian profile as an estimate of the background contribution arising from atom heating and from atom scattering halos resulting from the pumps. Then the self-correlation analysis is performed on the background-subtracted images. Due to imperfect subtraction, negative values appear in parts of the correlation. Note that since we are only interested in the correlations between Bragg peaks---all positive valued---the negative values do not affect the results. This analysis is repeated for each value of probe detuning $\omega$ and $k_\perp$ to form the experimental $\omega(k_\perp)$ dispersion curve shown in Fig.~\ref{fig4}.  Due to the sensitivity to atom number fluctuations in the correlation versus $\omega$ spectroscopy data, we perform bootstrap sampling to obtain a more reliable error estimate for the data points comprising $\omega(k_\perp)$.

\hl{\section{Cavity-mediated interaction in a confocal cavity}}
\label{sec:cavity_mediated_int}

In this section, we present the cavity-mediated interaction in a near-confocal cavity and show how the nonlocal term may be canceled through double pumping, leaving only a local interaction that allows for phonon excitations.  This follows the treatment in our Ref.~\cite{Guo2019eab}. To model an imperfectly degenerate cavity, as a first-order approximation, we take mode detunings $\Delta_\mu = \Delta_C + n_{\mu} \epsilon$. Here, $\Delta_C$ is the pump detuning to the reference mode (typically the peak of the mode spectrum; see, e.g.,  Fig.~\ref{pumping}) and $\epsilon$ is the residual mode splitting.  In the case of large cavity detuning, the cavity-mediated interaction is well-described by $\epsilon = 5$ MHz~\cite{Vaidya2017tpa}. For notational ease, we index the transverse cavity modes with the single variable $\mu$, rather than separate indices $l$ and $m$ for the Hermite-Gauss functions along the $\hat{x}$ and $\hat{y}$ directions, respectively.  We thus use $\mu = \{l,m\}$ to label the transverse electromagnetic mode function TEM$_{\mu}$ $\equiv$ TEM$_{l,m}$,  and we define the total mode family index $n_\mu = l + m$. As in the main text, we consider a scheme with two pumps, each of which is nearly resonant with a confocal resonance but which are separated by one FSR.  We index these confocal resonances by longitudinal mode numbers $Q_a$ and $Q_b$, such that $|Q_a-Q_b|=1$. 

We start from the general form of the cavity-mediated interaction in a confocal cavity~\cite{Guo2019eab} with only a single pump before generalizing to the case of two pumps. The interaction is
\bea\label{eqn:D3D}
\mathcal{D}_{3\text{D}}^\pm(\mbf{r},\mbf{r}^\prime,z,z^\prime) = \sum_{\mu}\frac{\Phi_\mu(\mbf{r},z)\Phi_\mu(\mbf{r}^\prime,z^\prime)}{1+\tilde\epsilon n_\mu}\mathcal{S}_\mu^\pm,
\eea
where $\tilde\epsilon=\epsilon/\Delta_C$, and we have neglected terms proportional to $\kappa/\Delta_C\ll1$ for simplicity. The terms $\mathcal{S}_\mu^\pm=[1\pm(-1)^{n_\mu}]/2$ enforce the parity symmetry of confocal cavities, which requires that only even or odd modes become degenerate at a single frequency. This symmetry results in degenerate resonances of even and odd modes separated by half of the FSR. Both pump frequencies are set near to even resonances; thus, the odd resonances are far-detuned by approximately 7.5~GHz and do not contribute to the interaction.

The full cavity mode functions appearing in the interaction are
\bea
\Phi_\mu(\mbf{r},z)=\Xi_\mu(\mbf{r})\cos\left[k_r\left(z+\frac{r^2}{2R(z)}\right)-\theta_\mu(z)\right].
\eea
Above, $\Xi_\mu(\mbf{r})$ are the Hermite-Gauss mode functions and $R(z)=z+z_R^2/z$ is the radius of curvature of the cavity phase fronts, where $z_R$ is the Rayleigh range. The term $\theta_\mu(z)$ accounts for the Gouy phase shift---the fact that in a confocal cavity,  different transverse modes, though degenerate, have different longitudinal phase variation. This is given by
\begin{align}
  \theta_{\mu}(z) &= \vartheta(z) + n_\mu [\vartheta(L/2) + \vartheta(z)] -\Theta, \label{gouy}\\
  \vartheta(z) &= \mrm{arctan} \left( \frac{z}{z_R} \right).
\end{align}
The phase offset is $\Theta=(Q+1)\pi/2$ where $Q$ is the longitudinal mode index.  The form of this phase offset is fixed by the boundary condition that the light field vanishes at the two mirrors. From these equations, it is seen that when focusing on the behavior near the cavity midplane, $z=0$, there is a $\pi/2$ phase shift between two degenerate resonances separated by one cavity FSR; see Ref.~\cite{Guo2019eab} for more details. We suppress the dependence of $\mathcal{D}_{3\text{D}}^\pm$ on $Q$ except when necessary for clarity.   

Evaluation of the cavity-mediated interaction requires generalization of the known Green's function of the harmonic oscillator to account for mode dispersion and parity symmetry in a confocal cavity. The starting point is the Mehler kernel, which gives the harmonic oscillator Green's function
\begin{equation}
G(\mbf{r},\mbf{r}^\prime,\varphi)
=
\sum_{\mu}\Xi_{\mu}(\mbf{r})\Xi_{\mu}(\mbf{r}^\prime) e^{-n_\mu \varphi} 
= \frac{e^{\varphi}}{2 \pi \sinh(\varphi)}
\exp\left[- \frac{(\mbf{r}-\mbf{r}^\prime)^2/w_0^2}{2 \tanh(\varphi/2)} -
\frac{(\mbf{r}+\mbf{r}^\prime)^2/w_0^2}{2 \coth(\varphi/2)}
\right],
\label{harmonicgreen}
\end{equation}
where $w_0$ is the waist of the TEM$_{0,0}$ mode, $35$~$\mu$m in our case, and $\varphi$ is any complex argument. To account for the $\mu$-dependent denominator in Eqn.~\eqref{eqn:D3D} that describes the dispersion of cavity modes, we define a modified Green's function as
\bea 
\mathcal{G}(\mbf{r},\mbf{r}^\prime,\varphi)=\sum_\mu\frac{\Xi_\mu(\mbf{r})\Xi_\mu(\mbf{r}^\prime)}{1+\tilde\epsilon n_\mu }e^{-n_\mu \varphi}
=\int_0^\infty d\lambda e^{-\lambda}G(\mbf{r},\mbf{r}^\prime,\varphi+\tilde\epsilon\lambda).
\eea
The modified Green's function can be computed analytically in the limit $\tilde\epsilon\ll 1$, which will be performed below. Parity symmetry is then accounted for by defining symmetrized and antisymmetrized versions of the modified Green's function as
\be
\mathcal{G}^\pm(\mbf{r},\mbf{r}^\prime,\varphi) = \mathcal{G}(\mbf{r},\mbf{r}^\prime,\varphi)\pm \mathcal{G}(\mbf{r},-\mbf{r}^\prime,\varphi),
\ee
where the $+(-)$ sign is taken for cavities of even (odd) parity. The confocal interaction may now be written in terms of $\mathcal{G}^\pm$ by expressing $S_\mu^\pm$ in terms of exponential functions. Using that $z_R=L/2$ and thus $\vartheta(L/2)=\pi/4$ in a confocal cavity, we find
\bea
\mathcal{D}_{3\text{D}}^\pm(\mbf{r},\mbf{r}^\prime,z,z^\prime) &=& \frac{1}{4}\sum_{\tau,\tau^\prime=\pm1} \mathcal{G}^\pm\big(\mbf{r},\mbf{r}^\prime, -i\pi(\tau+\tau^\prime)/4 -i[\tau\vartheta(z)+\tau^\prime\vartheta(z^\prime)] \big) \nonumber \\
&&\qquad\quad\quad\times \exp\big[ik_r(\tau z+\tau^\prime z^\prime)-i[\tau\vartheta(z)+\tau^\prime\vartheta(z^\prime)]+i\pi(Q+1)(\tau+\tau^\prime)/2\big].
\eea
We now specialise to the case of an even confocal cavity and consider atoms located near the center of the cavity, so that $|z|\ll z_R$. In this case, $\vartheta(z)$ is negligible and the interaction becomes
\bea
\mathcal{D}_{3\text{D}}^+(\mbf{r},\mbf{r}^\prime,z,z^\prime) =\quad &\frac{1}{2} \left[\mathcal{G}^+(\mbf{r},\mbf{r}^\prime,0)+(-1)^{Q+1}\mathcal{G}^+(\mbf{r},\mbf{r}^\prime,i\pi/2)\right]\cos k_rz \cos k_rz^\prime \nonumber \\
 + &\frac{1}{2} \left[\mathcal{G}^+(\mbf{r},\mbf{r}^\prime,0)-(-1)^{Q+1}\mathcal{G}^+(\mbf{r},\mbf{r}^\prime,i\pi/2)\right]\sin k_rz \sin k_rz^\prime,
 \label{eqn:D3Dsigns}
\eea
where we have used that $\mathcal{G}^\pm(\mbf{r},\mbf{r}^\prime,-i\pi/2)=\pm\mathcal{G}^\pm(\mbf{r},\mbf{r}^\prime,i\pi/2)$. This is the explicit form of Eq.~\eqref{eqn:IntA} and \eqref{eqn:IntB} in the main text. The terms $\mathcal{G}^+(\mbf{r},\mbf{r}^\prime,0)$ correspond to local and mirror image interactions while $\mathcal{G}^+(\mbf{r},\mbf{r}^\prime,i\pi/2)$ corresponds to the nonlocal interaction.

We now consider the form of the $z-$independent terms in $\mathcal{D}_{3\text{D}}^+$ more explicitly to connect to Eq.~\eqref{eqn:intStrength} before adding the second pump. The transverse dependence of $\mathcal{D}_{3\text{D}}^+$ may be written as
\bea 
\mathcal{D}^+(\mbf{r},\mbf{r}^\prime) = \mathcal{D}_{\text{loc}}(\mbf{r},\mbf{r}^\prime) + \mathcal{D}_{\text{loc}}(\mbf{r},-\mbf{r}^\prime) \pm \mathcal{D}_{\text{non}}(\mbf{r},\mbf{r}^\prime),
\eea 
where $\mathcal{D}_{\text{loc}}(\mbf{r},\mbf{r}^\prime)$ is the local interaction, $\mathcal{D}_{\text{loc}}(\mbf{r},-\mbf{r}^\prime)$ is the mirror image term, and $\mathcal{D}_{\text{non}}(\mbf{r},\mbf{r}^\prime)$ is the nonlocal interaction that must be canceled to obtain the $U(1)$ symmetry. The local and nonlocal interactions may be accurately approximated when $\tilde\epsilon\ll 1$. Under this condition, the nonlocal interaction becomes~\cite{Vaidya2017tpa} 
\bea 
\mathcal{D}_{\text{non}}(\mbf{r},\mbf{r}^\prime)=\frac{\cos\left(\dfrac{\mbf{r}\cdot\mbf{r}^\prime}{w_0^2/2}\right)}{4\pi\big[1+\tilde\epsilon(\mbf{r}^2+{\mbf{r}^\prime}^2)/w_0^2\big]}.
\eea
A simplified version of this nonlocal interaction is presented in Eq.~\eqref{eqn:intStrength}.
The local interaction takes the form
\bea 
\mathcal{D}_{\text{loc}}(\mbf{r},\mbf{r}^\prime)= \frac{1}{2 \pi \tilde{\epsilon}} K_0 \left( \frac{2}{\sqrt{\tilde{\epsilon}}} \left| \frac{\mbf{r} - \mbf{r}^\prime}{w_0} \right| \sqrt{1  + \frac{\tilde{\epsilon}}{4}\left(\frac{\mbf{r} + \mbf{r}^\prime}{w_0}\right)^2}\right),
\eea
where $K_0$ is the modified Bessel function of the second kind. Assuming that the spatial extent of atomic cloud is smaller than $2w_0/\sqrt{\tilde{\epsilon}}\approx 220$~$\mu$m, which is true for our condensate length of $\sim$93~$\mu$m, the local interaction may be approximated as 
\begin{align}\label{localinteraction}
\mathcal{D}_{\text{loc}} (\mbf{r},\mbf{r}^\prime) &\approx \frac{1}{2 \pi \tilde{\epsilon}} K_0 \left( \frac{2}{\sqrt{\tilde{\epsilon}}} \left| \frac{\mbf{r} - \mbf{r}^\prime}{w_0} \right| \sqrt{1 + \tilde{\epsilon} \left(\frac{\mbf{r}_{\mathrm{cm}}}{w_0}\right)^2}\right) \nonumber \\
&\equiv K(|\mbf{r} - \mbf{r}^\prime|/\xi), 
\end{align}
where $\mathbf{r}_{\mathrm{cm}} = (\text{49~}\mu\mathrm{m},\text{35~}\mu\mathrm{m})$ is the centre of mass coordinate of the atomic cloud and the length scale associated with local interaction is 
\be
\xi \equiv \frac{w_0 \sqrt{\tilde{\epsilon}}}{2 \sqrt{1 + \tilde{\epsilon} (\mbf{r}_{\mathrm{cm}}/w_0)^2}}.
\ee
The function $K(|\delta\mbf{r}|/\xi)$ falls off as $\exp(-|\delta\mbf{r}|/\xi)/\sqrt{|\delta\mbf{r}|/\xi}$ for large $|\mbf{r}-\mbf{r}^\prime|$, as presented in Eq.~\eqref{eqn:intStrength}. In the limit that $\sqrt{\tilde{\epsilon}}{r}_{\mathrm{cm}}/w_0 \ll 1$, we have that $\xi \approx w_0 \sqrt{\tilde{\epsilon}}/2$. For our system, this local interaction range is around 5~$\mu$m at the $\Delta_C$ employed for the spectroscopy data taken here~\cite{Vaidya2017tpa,Kroeze2021preprint}.
By placing the atoms in a single half-plane of the cavity, we can ensure that the local mirror-image  term is irrelevant and can be neglected.

To achieve the $U(1)$ symmetry, the nonlocal interaction must be canceled by pumping on two confocal resonances separated in longitudinal mode number by $|Q_a-Q_b|=1$ with matched Raman coupling rates. In this case, the $Q$-dependent signs in Eq.~\eqref{eqn:D3Dsigns} mean the nonlocal terms from each resonance cancel, and the total interaction is
\bea
\mathcal{D}_{\text{double}}^+(\mbf{r},\mbf{r}^\prime,z,z^\prime) &=& \mathcal{D}_{3\text{D}}^+(\mbf{r},\mbf{r}^\prime,z,z^\prime;Q_a) + \mathcal{D}_{3\text{D}}^+(\mbf{r},\mbf{r}^\prime,z,z^\prime;Q_b)  \nonumber \\
&=& K(|\mbf{r}-\mbf{r}^\prime|/\xi)\cos[k_r(z-z^\prime)].
\eea
The above equation represents the explicit form of Eq.~\eqref{UmmTotal} in the main text, demonstrating both the $U(1)$ translation symmetry in the $z$-direction and local interaction in the transverse plane achieved through double pumping.

\section{Self-organisation in a double-pumped confocal cavity}

To derive a theory for obtaining the dispersion-relation curves plotted in the figures of the main text, we start with the Hamiltonian describing atoms coupled to two degenerate resonances of a confocal cavity under the transverse double-pump scheme:
\begin{align}
\label{eq:Hamiltonian}
H=& -\sum_{\mu}\Delta_{\mu} \hat{a}^\dagger_{\mu} \hat{a}^{}_{\mu} -\sum_{\mu}\Delta_{\mu} \hat{b}^\dagger_{\mu} \hat{b}^{}_{\mu} \nonumber \\
&+ \int d^3\mbf{x}
\hat{\Psi}^\dagger(\mbf{x})\left(-\frac{\nabla^2}{2m}+ V(\mbf{x}) 
+U|\hat{\Psi}(\mbf{x})|^2\right)\hat{\Psi}(\mbf{x})\nonumber \\
&+ \int d^3\mbf{x}
\hat{\Psi}^\dagger(\mbf{x})\left(\frac{|\hat{\phi}_a|^2}{\Delta^a_A} + \frac{|\hat{\phi}_b|^2}{\Delta^b_A}\right)
\hat{\Psi}(\mbf{x}),
\end{align}
where $N$ is the number of atoms and $\Delta^a_A$ and $\Delta^b_A$ are the atomic detuning of the two pumps.  The terms in the first line are the cavity field energies in a frame rotating at one FSR.  The terms in the second line are the atomic kinetic energy, potential energy from a trap $V(x)$, and contact interaction of strength $U$ (note this is  separate from the cavity mediated atom-atom interaction discussed in the main text). The third line contains the interaction terms between the BEC and the two cavity mode families. The matter wave field is denoted by $\hat{\Psi}(\mbf{x})$, while the light fields are $\hat{\phi}_a$ and $\hat{\phi}_b$, which contain both the standing-wave transverse pump and a sum over all cavity modes with transverse and longitudinal spatial dependence.  In writing the above, we have set $\hbar=1$, and we will continue to do so throughout this and following sections.

The light field contains two transverse pumps with strength $\Omega_a$ and $\Omega_b$ coupled to two degenerate families of cavity modes separated by one FSR.    The total fields are
\begin{align}
\hat{\phi}_a(\mbf{r}) &= \Omega_a \cos(k_rx) \nonumber \\
&+ g_0\sum_{\mu} \hat{a}_{\mu}
\Xi_\mu(\mbf{r})\cos{\left[k_r\left(z+\frac{r^2}{2R(z)}\right)-\theta_{\mu}(z)\right]}, \nonumber \\
\hat{\phi}_b(\mbf{r}) &= \Omega_b \cos(k_rx)  \nonumber \\ &+g_0\sum_{\mu} \hat{b}_{\mu}
\Xi_\mu(\mbf{r})\cos{\left[k_r\left(z+\frac{r^2}{2R(z)}\right)-\theta_{\mu}(z) + \pi/2\right]},
\label{totalLightField}
\end{align}
where $\Xi_{\mu}(\mbf{r})$ is the spatial profile of a Hermite-Gauss mode of the cavity and the summation runs only over even or odd modes. The form of the light field
results in a spatially varying single-photon Rabi frequency $g_0 \Xi_{\mu} (\mbf{r})/\Xi_{0,0}(0)$. The term $\theta_\mu(z)$ is again given by Eq.~\eqref{gouy} using the longitudinal mode number $Q_a$ of the first resonance.

\subsection{Derivation of equations of motion}

\subsubsection{Effective Hamiltonian}

In the main text, we discussed the effective atom-atom interaction mediated by the cavity,  specialising to atoms at the cavity midplane, $z=0$.  In this section, we discuss the cavity-mediated interaction more fully.  To do this, it is clearer to consider this as an interaction between atomic density waves.  We therefore expand the atomic wavefunction as
\hl{
\be
\hat{\Psi} = Z(z - z_0) \left\{ \hat{\psi}_0 (\mbf{r}) \mu_0 (k_r x) +  \sqrt{2}[\hat{\psi}_c (\mbf{r}) \cos(k_r z + \delta) + \hat{\psi}_s (\mbf{r}) \sin(k_r z + \delta) ] \mu_1(k_r x) \right\}e^{- i \mathcal{E}_0 t},
\label{atomexpansion}
\ee}%
where $\mu_n(\phi)$ are the $2 \pi$ periodic eigenfunctions of the Mathieu equation, $\partial_\phi^2 \mu_n + [a_n - 2 \mathcal{Q} \cos(2 \phi)] \mu_n=0$,  with eigenvalues $a_n$ and the $\mu_n(\phi)$ describing wavefunctions in the pump lattice, \hl{and $\mathcal{E}_0$ is the chemical potential of the original condensate}. The dimensionless parameter 
$\mathcal{Q} = -\Omega_a^2/(4 \Delta^a_A \omega_r) - \Omega_b^2/(4 \Delta^b_A \omega_r)$ is the pump lattice depth in units of the recoil energy $\omega_r = k_r^2/2m_{\mrm{Rb87}}$.
The factor $Z(z)$ is the envelope function in $\hat{z}$; $\psi_0$ is the condensate wavefunction; \hl{$\psi_{c,s}$ are the envelope functions of the atomic density wave scattered into two (mutually out-of-phase)  profiles};  $\delta$ is a fixed phase offset that we will later choose for convenience.

Our aim will be to derive coupled equations for the \hl{condensate $\psi_0$, the density wave envelope functions $\psi_{c,s}$}, and the cavity light;  this will allow us to find the dispersion of the normal modes (DW polaritons), both below and above the DW polariton condensation threshold.
To consider the cavity fields, we may first focus on a single degenerate resonance, and then later combine the effects of both resonances. The linear-order light-matter coupling term in the Hamiltonian is
\be\label{atomwave}
H^{a}_{LM}=
\frac{g_0 \Omega}{\Delta^a_A} \int d^3 \mbf{x} |\Psi(\mbf{x})|^2 \sum_{\mu}\Xi_{\mu}(\mbf{r})(\hat{a}^{\dagger}_{\mu} + \hat{a}_{\mu}) \cos(k_r x) \cos[k_r z - \theta_\mu (z)].
\ee
We have assumed that the Rayleigh range is much larger than the BEC, enabling us to drop the $r^2/2R(z)$ term.  We next integrate-out the $z$ dependence because the dynamics of interest occur in the transverse plane. This can be done straightforwardly in the limit where we assume $Z(z-z_0)$ has a width $w_z$ and that $\lambda\ll w_z\ll z_R$. The first inequality allows us to drop any terms oscillating at wavevectors $k_r$ or $2k_r$ along $\hat z$; this imposes momentum conservation so that recoiling atoms receive momentum kicks given by the difference between the pump and cavity momenta.  The second condition means that we can evaluate the slowly varying phase terms as being effectively constant over the width of the gas:  $\theta_{\mu}(z) \simeq \theta_{\mu}(z_0)$. Similarly, we will drop the fast oscillating terms along $\hat{x}$. Using the expression for the atomic wavefunction in Eq.~\eqref{atomexpansion} and keeping terms up to linear order in \hl{$\hat{\psi}_{c,s}$}, the coupling term then becomes
\hl{
\be
H^{a}_{LM}=\eta_a \sqrt{2}  \int d \mbf{r} \sum_{\mu}\Xi_{\mu}(\mbf{r})(\hat{a}^{\dagger}_{\mu} + \hat{a}_{\mu})\left\{\hat{\psi}^{\dagger}_0(\mbf{r}) \left[\hat{\psi}^{}_{c}(\mbf{r})  +i\hat{\psi}^{}_{s}(\mbf{r})\right] e^{i (\delta + \theta_{\mu}(z_0))} + \hat{\psi}^{}_0(\mbf{r}) \left[\hat{\psi}^{\dagger}_{c}(\mbf{r})  +i\hat{\psi}^{\dagger}_{s}(\mbf{r})\right] e^{i (\delta + \theta_{\mu}(z_0))} + \text{H.c.} \right\},
\ee
}%
where $\eta_{a,b} \equiv O(\mathcal{Q}) g_0 \Omega_{a,b}/4\Delta^{a,b}_A$ is the two-photon coupling strength.  The factor $O(\mathcal{Q}) = \langle \mu_1(\phi) \mu_0(\phi) \cos(\phi) \rangle$ is the overlap of scattered atoms with the pump potential and the condensate in the pump lattice averaged over one lattice period $\phi \in [0,2 \pi]$. This overlap factor depends on the dimensionless pump lattice strength $\mathcal{Q}$ via the form of the Mathieu functions $\mu_n$ as defined above. \hl{We note that in the main text, we neglected this overlap factor for notational simplicity.}  To simplify the expression,  we choose $\delta = \Theta - \vartheta(z_0)$ and define $\theta_0 = \pi/4 + \mrm{arctan}(z_0/z_R)$.  We finally arrive at the following form of the interaction
\hl{
\be
H^{a}_{LM}=\eta_a \sqrt{2}  \int d \mbf{r} \sum_{\mu}\Xi_{\mu}(\mbf{r})(\hat{a}^{\dagger}_{\mu} + \hat{a}_{\mu})\left\{\hat{\psi}^{\dagger}_0(\mbf{r}) \left[\hat{\psi}^{}_{c}(\mbf{r})  +i\hat{\psi}^{}_{s}(\mbf{r})\right] e^{-i n_\mu \theta_0} + \hat{\psi}^{}_0(\mbf{r}) \left[\hat{\psi}^{\dagger}_{c}(\mbf{r})  +i\hat{\psi}^{\dagger}_{s}(\mbf{r})\right] e^{-i n_\mu \theta_0} + \text{H.c.} \right\}.
\ee}
The calculation for the degenerate resonance one FSR away (to the red detuning side)---the $\hat b_{\mu} + \hat b_{\mu}^\dagger$ modes---is identical to the above, except there is an additional $\pi/2$ phase shift in $\theta_{\mu}$ that shifts the longitudinal cavity profile. 

\hl{We now turn to the atomic Hamiltonian. The kinetic energy term is given by
\begin{align}
H_{KE} &= \int d^3\mbf{x} \hat{\Psi}^{\dagger}(\mbf{r})\left( - \frac{\nabla^2}{2 m } \right) \hat{\Psi} \nonumber \\
&= \int d\mbf{r} \hat{\psi}^{\dagger}_0(\mbf{r})\left[-\frac{\nabla^2}{2m} + \omega_r a_0 (\mathcal{Q}) \right]\hat{\psi}^{}_0(\mbf{r}) + \sum_{i=c,s} \hat{\psi}^{\dagger}_{i} \left[-\frac{\nabla^2}{2m} + \omega_r(1 + a_1(\mathcal{Q})) \right]\hat{\psi}_{i}
\end{align}
This expression shows the kinetic energy has two contributions, one from the spatial variation of the envelope functions, and one from Mathieu functions.  In writing the first contribution, we note that in our setup, the presence of a deep pump lattice can also change the dispersion along the pump direction.  This would mean that in general the dispersion becomes anisotropic in the $xy$ plane.  Since our experiments only probe variations along the $x$ direction, we can account for this modified dispersion by using $m$ as the effective mass of the atoms in  the band structure of the pump lattice, different from the bare atomic mass $m_{\mrm{Rb87}}$.}

\hl{The atomic interaction term becomes
\begin{align}
H_{\mrm{int}} &= \int d^3{\mbf{x}} U |\hat{\Psi}|^4 \nonumber \\
&= U_{2D}\int d\mbf{r} \Bigg\{ \langle \mu^4_0 \rangle |\hat{\psi}^{}_0(\mbf{r})|^4 + \langle \mu^2_0 \mu^2_1 \rangle \left[ \hat{\psi}^{\dagger 2}_0(\mbf{r}) \left(\hat{\psi}^{2}_c(\mbf{r}) + \hat{\psi}^{2}_s(\mbf{r})\right) + \text{H.c.} \right] + \langle \mu^2_0 \mu^2_1 \rangle 4 |\hat{\psi}^{}_0(\mbf{r})|^2 \left( |\hat{\psi}^{}_c(\mbf{r})|^2 + |\hat{\psi}^{}_s(\mbf{r})|^2\right) \nonumber \\
& + \langle \mu^4_1 \rangle \left[ \frac{3}{2} \left( |\hat{\psi}^{}_c(\mbf{r})|^4 + |\hat{\psi}^{}_s(\mbf{r})|^4 \right) + \frac{1}{2} \left( 4|\hat{\psi}^{}_c(\mbf{r})|^2 |\hat{\psi}^{}_s(\mbf{r})|^2 + \hat{\psi}^{\dagger 2}_c(\mbf{r})\hat{\psi}^{2}_s(\mbf{r}) + \hat{\psi}^{\dagger 2}_s(\mbf{r})\hat{\psi}^{2}_c(\mbf{r}) \right) \right]\Bigg\},
\end{align}
where $U_{2D}$ is the effective interaction in 2D, which, due to the harmonic confinement of the atoms along the $z$ direction, is given by the 3D interaction renormalized by the harmonic oscillator length $w_z = \sqrt{\pi/(m_{\mrm{Rb87}} \omega_z)}$, i.e. $U_{2D} = U/w_z$. Given our atomic density $\rho$, we calculated $U_{2D}\rho/\hbar = 2\pi \times 254 $ Hz, which is roughly $\sim$5\% of the kinetic energy of the scattered atoms. For simplicity in later derivations, we define the following spatially averaged atomic interactions
\be
\label{eq:defUeff}
U_{00} = U_{2D} \langle \mu^4_0 \rangle ,\quad U_{01} = U_{2D} \langle \mu^2_0 \mu^2_1 \rangle,\quad U_{11} = U_{2D} \langle \mu^4_1 \rangle,
\ee
which thus depend implicitly on the pump parameter $\mathcal{Q}$.
} 

\hl{Putting these together, we can now rewrite the full Hamiltonian up to linear order in light-matter coupling as
\begin{align}
H =& -\sum_{\mu}\Delta_{\mu} \hat{a}^\dagger_{\mu} \hat{a}^{}_{\mu} -\sum_{\mu}\Delta_{\mu} \hat{b}^\dagger_{\mu} \hat{b}^{}_{\mu} \nonumber \\
&+\int d\mbf{r} \hat{\psi}_{0}^{\dagger}(\mbf{r}) \left[-\frac{\nabla^2}{2m} + \omega_r a_0(\mathcal{Q})+ V(\mbf{r}) 
+U_{00} |\hat{\psi}^{}_0(\mbf{r})|^2  \right]\hat{\psi}^{}_{0} \nonumber \\
&+ \sum_{i=c,s} \int d\mbf{r} \Bigg\{ \hat{\psi}_{i}^{\dagger}(\mbf{r}) \left[-\frac{\nabla^2}{2m}+ \omega_r(1+ a_1(\mathcal{Q})) + V(\mbf{r}) 
+ 4 U_{01} |\hat{\psi}^{}_0(\mbf{r})|^2 \right] \hat{\psi}^{}_{i}(\mbf{r}) + U_{01} \left[ \hat{\psi}^{\dagger 2}_0(\mbf{r})\hat{\psi}^{2}_i(\mbf{r}) + \text{H.c.} \right] + \frac{3}{2}U_{11} |\hat{\psi}^{}_i|^4 \Bigg\} \nonumber \\ 
&+ \int d\mbf{r} \frac{U_{11}}{2} \left( 4|\hat{\psi}^{}_c(\mbf{r})|^2 |\hat{\psi}^{}_s(\mbf{r})|^2 + \hat{\psi}^{\dagger 2}_c(\mbf{r})\hat{\psi}^{2}_s(\mbf{r}) + \hat{\psi}^{\dagger 2}_s(\mbf{r})\hat{\psi}^{2}_c(\mbf{r}) \right) \nonumber \\
&+ \eta_a \sqrt{2}  \int d \mbf{r} \sum_{\mu}\Xi_{\mu}(\mbf{r})(\hat{a}^{\dagger}_{\mu} + \hat{a}_{\mu})\left\{\hat{\psi}^{\dagger}_0(\mbf{r}) \left[\hat{\psi}^{}_{c}(\mbf{r})  +i\hat{\psi}^{}_{s}(\mbf{r})\right] e^{-i n_\mu \theta_0} + \hat{\psi}^{}_0(\mbf{r}) \left[\hat{\psi}^{\dagger}_{c}(\mbf{r})  +i\hat{\psi}^{\dagger}_{s}(\mbf{r})\right] e^{-i n_\mu \theta_0} + \text{H.c.} \right\}  \nonumber \\
&+ \eta_b \sqrt{2}  \int d \mbf{r} \sum_{\mu}\Xi_{\mu}(\mbf{r})(\hat{b}^{\dagger}_{\mu} + \hat{b}_{\mu})\left\{i\hat{\psi}^{\dagger}_0(\mbf{r}) \left[\hat{\psi}^{}_{c}(\mbf{r})  +i\hat{\psi}^{}_{s}(\mbf{r})\right] e^{-i n_\mu \theta_0} + i\hat{\psi}^{}_0(\mbf{r}) \left[\hat{\psi}^{\dagger}_{c}(\mbf{r})  +i\hat{\psi}^{\dagger}_{s}(\mbf{r})\right] e^{-i n_\mu \theta_0} + \text{H.c.} \right\}.
\label{eq:lin-ham}
\end{align}}
Note that the additional factor of $i$ in the last line of Eq.~\eqref{eq:lin-ham} is due to the aforementioned $\pi/2$ shift for $\hat b_\mu$ modes. We again take $\Delta_\mu = \Delta_C + n_{\mu} \epsilon$ to model an imperfect confocal cavity.

\hl{The model in Eq.~\eqref{eq:lin-ham} is very closely related to those used to model exciton-polariton condensates in semiconductor microcavities, forming ``quantum fluids of light''~\cite{Carusotto2013qfo}.  As we will see below,  our DW polaritons interact both due to the atomic nonlinear term $U|\Psi(\mbf{r})|^4$, as well as due to the saturability of the matter-light coupling term.  This is directly analogous to the exciton-polariton system.  Some important differences are however worth noting.  Firstly, in our case, the cavity detuning  $\Delta_C$ is generally large---much larger than the DW energy ${\sim}\omega_r$, so we are far from resonance.  This separation of energy scales also allows us to approximately describe the system by adiabatically eliminating the photons (as done in the main text) to produce an effective interaction. This separation of energy scales also means that at low pumping powers, the normal modes of our system are initially density waves and photons. These only become mixed when the pump-induced coupling $\eta_{a,b}$ is sufficiently large.  Secondly, as discussed above, the optical modes in our case are those of a confocal cavity, rather than the planar cavities typically studied for exciton-polaritons. Finally, a critical difference from exciton-polaritons is that in our case there is no incoherent pumping required to reach condensation.  Instead, the Raman pumping scheme means the effective ``synthetic'' light-matter coupling $\eta_{a,b}$ can be sufficient to induce the superradiance phase transition, despite the presence of cavity losses~\cite{Kirton2018itt,Mivehvar2021cqw}.}

\subsubsection{Continuous light field representation in confocal limit}

We now focus on the case of two confocal degenerate resonances that contain only the even transverse modes, i.e., $n_\mu~\mathrm{mod}~2 = 0$, \hl{and assume balanced pumping $\eta_a=\eta_b\equiv\eta$}. To account for the infinite number of transverse modes in a more tractable manner, we will find it useful to define the following cavity operators:
\begin{align}
\hat{\mathcal{A}}(\mathbf{r}) &= \frac{1}{\sqrt{\mathcal{N}_a}}\sum_{\mu} \frac{\Xi_{\mu}(\mbf{r})}{w_0/\sqrt{2}} \left[ \hl{\hat{a}_{\mu} \cos(n_{\mu} \theta_0) + \hat{b}_{\mu} \sin(n_{\mu} \theta_0)} \right] \mathcal{S}^{+}_{\mu}, \nonumber \\
\hat{\mathcal{B}}(\mathbf{r}) &= \frac{1}{\sqrt{\mathcal{N}_b}}\sum_{\mu} \frac{\Xi_{\mu}(\mbf{r})}{w_0/\sqrt{2}} \left[ \hl{-\hat{a}_{\mu} \sin(n_{\mu} \theta_0) + \hat{b}_{\mu} \cos(n_{\mu} \theta_0)} \right]\mathcal{S}^{+}_{\mu},
\label{light_reduce}
\end{align}
where $\mathcal{N}_{a,b}$ are normalisation factors to guarantee bosonic commutation relations and the factor
\be
\mathcal{S}^{+}_{\mu} = \frac{1}{2} [1 + (-1)^{n_\mu}]
\ee
is chosen to cancel the odd modes in a degenerate confocal resonance such that the summation can be carried over all transverse modes. Computing the commutation relations, we find that
\begin{align}
    [\hat{\mathcal{A}}(\mbf{r}), \hat{\mathcal{A}}^{\dagger}(\mbf{r}^\prime) ] &= \frac{1}{\mathcal{N}_a}\sum_{\mu} \frac{\Xi_{\mu} (\mbf{r})}{w_0/\sqrt{2}} \frac{\Xi_{\mu} (\mbf{r}^\prime)}{w_0/\sqrt{2}}\mathcal{S}^{+}_{\mu} = \frac{1}{2 \mathcal{N}_a} [\delta(\mbf{r} - \mbf{r}^\prime)+\delta(\mbf{r} + \mbf{r}^\prime)], \nonumber \\
    [\hat{\mathcal{B}}(\mbf{r}), \hat{\mathcal{B}}^{\dagger}(\mbf{r}^\prime) ] &= \frac{1}{\mathcal{N}_b}\sum_{\mu} \frac{\Xi_{\mu} (\mbf{r})}{w_0/\sqrt{2}} \frac{\Xi_{\mu} (\mbf{r}^\prime)}{w_0/\sqrt{2}}\mathcal{S}^{+}_{\mu} = \frac{1}{2 \mathcal{N}_b} [\delta(\mbf{r} - \mbf{r}^\prime)+\delta(\mbf{r} + \mbf{r}^\prime)],
\end{align}
where the normalisation condition of the Hermite-Gauss mode function is given by
\be
\int \frac{d \mbf{r}}{w^2_0/2} \Xi_{\mu}(\mbf{r}) \Xi_{\mu}(\mbf{r}) = 1.
\ee
The appearance of the mirror-image term $\delta(\mbf{r} + \mbf{r}^\prime)$ is due to the fact that the summation is restricted to all modes with even spatial symmetry. We note that this term does not play a role in the atom-cavity interaction because we place the BEC away from the cavity centre so that no atoms exist at the mirror image position.  Consequently, the same normalisation $\mathcal{N}_{a,b} =  1/2$ satisfies the bosonic commutation relation for both $\hat{\mathcal{A}}$ and $\hat{\mathcal{B}}$. Using orthonormality of the Hermite-Gauss mode functions $\Xi_{\mu}$, the original boson modes may be rewritten as
\begin{align}
\hat{a}_{\mu} &= \sqrt{\frac{1}{2}} \int d\mbf{r} \frac{\Xi_{\mu}(\mbf{r})}{w_0/\sqrt{2}}  \left[  \hl{\hat{\mathcal{A}} (\mbf{r}) \cos(n_{\mu} \theta_0) - \hat{\mathcal{B}}(\mbf{r}) \sin(n_{\mu} \theta_0)} \right], \nonumber \\
\hat{b}_{\mu} &= \sqrt{\frac{1}{2}} \int d\mbf{r} \frac{\Xi_{\mu}(\mbf{r})}{w_0/\sqrt{2}}  \left[  \hl{ \hat{\mathcal{A}} (\mbf{r}) \sin(n_{\mu} \theta_0) + \hat{\mathcal{B}}(\mbf{r}) \cos(n_{\mu} \theta_0) }\right].
\end{align}

Employing the $\hat{\mathcal{A}}$, $\hat{\mathcal{B}}$ basis,  we can now rewrite the original Hamiltonian in terms of an inverse Green's function $\mathcal{D}^{-1}(\mbf{r},\mbf{r}^\prime)$: %\begin{widetext}
\hl{
\begin{align}
H =& -\Delta_C \int \frac{d\mbf{r} d \mbf{r}^\prime}{w^2_0/2} [\hat{\mathcal{A}}^{\dagger}(\mbf{r})\mathcal{D}^{-1}(\mbf{r},\mbf{r}^\prime)\hat{\mathcal{A}}(\mbf{r}^\prime)+\hat{\mathcal{B}}^{\dagger}(\mbf{r})\mathcal{D}^{-1}(\mbf{r},\mbf{r}^\prime)\hat{\mathcal{B}}(\mbf{r}^\prime)]  \nonumber \\ 
&+\int d\mbf{r} \hat{\psi}_{0}^{\dagger}(\mbf{r}) \left[-\frac{\nabla^2}{2m} + \omega_r a_0(\mathcal{Q})+ V(\mbf{r}) 
+U_{00} |\hat{\psi}^{}_0(\mbf{r})|^2 \right]\hat{\psi}^{}_{0} \nonumber \\
&+ \sum_{i=c,s} \int d\mbf{r} \Bigg\{ \hat{\psi}_{i}^{\dagger}(\mbf{r}) \left[-\frac{\nabla^2}{2m}+ \omega_r(1+ a_1(\mathcal{Q})) + V(\mbf{r}) 
+ 4 U_{01} |\hat{\psi}^{}_0(\mbf{r})|^2 \right] \hat{\psi}^{}_{i}(\mbf{r}) + U_{01} \left[ \hat{\psi}^{\dagger 2}_0(\mbf{r})\hat{\psi}^{2}_i(\mbf{r}) + \text{H.c.} \right] + \frac{3}{2}U_{11} |\hat{\psi}^{}_i|^4 \Bigg\} \nonumber \\ 
&+ \int d\mbf{r} \frac{U_{11}}{2} \left( 4|\hat{\psi}^{}_c(\mbf{r})|^2 |\hat{\psi}^{}_s(\mbf{r})|^2 + \hat{\psi}^{\dagger 2}_c(\mbf{r})\hat{\psi}^{2}_s(\mbf{r}) + \hat{\psi}^{\dagger 2}_s(\mbf{r})\hat{\psi}^{2}_c(\mbf{r}) \right) \nonumber \\
&+ 2\eta \int  d \mbf{r} \frac{w_0}{\sqrt{2}} [\hat{\psi}_{c}^{\dagger}(\mbf{r})\hat{\psi}^{}_0 (\mbf{r}) + \hat{\psi}_{c}^{}(\mbf{r})\hat{\psi}^{\dagger}_0 (\mbf{r})][\hat{\mathcal{A}}^{\dagger}(\mbf{r})+\hat{\mathcal{A}}(\mbf{r})] - 2\eta \int d \mbf{r}\frac{w_0}{\sqrt{2}} [\hat{\psi}_{s}^{\dagger}(\mbf{r})\hat{\psi}^{}_0(\mbf{r}) + \hat{\psi}_{s}^{}(\mbf{r})\hat{\psi}^{\dagger}_0(\mbf{r})][\hat{\mathcal{B}}^{\dagger}(\mbf{r})+\hat{\mathcal{B}}(\mbf{r})],
\label{eq:realspaceH}
\end{align}
}
%\end{widetext}
where
\be
\mathcal{D}^{-1}(\mbf{r},\mbf{r}^\prime) = \sum_{\mu} (1 + n_\mu \tilde{\epsilon}) \Xi_{\mu}(\mbf{r}) \Xi_{\mu} (\mbf{r}^\prime),
\ee
and $\tilde{\epsilon} \equiv \epsilon/\Delta_C$. Here, the summation is over all transverse modes with both even and odd spatial symmetry since we have dropped the mirror-image term in the commutation relation for the cavity field operators $\hat{\mathcal{A}}$ and $\hat{\mathcal{B}}$. 

\hl{We note that this equation contains two sources of nonlinearity for the field $\psi_{c,s}$. As we will show later in the mean-field equation of motion, in addition to the atomic contact interaction $U$, there is nonlinearity implicit in the form of the atom-light coupling from the requirement of local atom number conservation for a uniform BEC, i.e., $|\psi_0|^2 + |\psi_c|^2 + |\psi_s|^2 = \rho$, where $\rho$ is the total atom density.}

Equation~\eqref{eq:realspaceH}, along with the definition of  $\mathcal{D}^{-1}(\mbf{r},\mbf{r}^\prime)$,
\hl{(which is the inverse of the Green's function defined in Sec.~\ref{sec:cavity_mediated_int})}, provides a general description of the cavity tuned near to a confocal point, but with a nonzero $\epsilon$.  We now seek the DW polariton dispersion relation using a momentum space description of the translation invariant interaction $K(|\mbf{r}-\mbf{r}^\prime|/\xi)$ derived in Eq.~\eqref{localinteraction}.
From the form of Eq.~\eqref{localinteraction}, we find that the approximate Green's function is diagonal in momentum space:
\be
\mathcal{D}(\mbf{k}) = \frac{1}{1 + k^2 /\zeta^2},
\ee
where $\zeta = 1/\xi$ is the characteristic momentum scale. Therefore, the dispersion of the cavity field $\hat{\mathcal{A}}$, $\hat{\mathcal{B}}$ in the small-$\tilde{\epsilon}$ limit is
\be
\mathcal{D}^{-1}(\mbf{k}) = 1 + k^2 /\zeta^2.
\ee
We note that the well-defined nature of the momentum peaks evident in our spectroscopy  experiments---indicating that momentum is a good quantum number---support the assumptions made above.

\subsubsection{Dissipative equations of motion}

Using the above translational invariance, we can now write the mean-field equations of motion for the cavity modes and atoms:
\hl{\begin{align}\label{EOM}
 i \partial_t \psi_{0} &= \left[ - \frac{\nabla^2}{2m} + \omega_r a_0(\mathcal{Q}) +  2U_{00} |\psi^{}_0|^2 -\mathcal{E}_0 \right] \psi_0 + U_{01}\left[4(|\psi_c|^2 + |\psi_s|^2) \psi_0 + 2 \psi^{\dagger}_0 (\psi^2_c + \psi^2_s) \right] \nonumber \\
 &+ 2 \eta \frac{ w_0}{\sqrt{2}}\left[(\mathcal{A}^{\ast} + \mathcal{A})\psi_c -(\mathcal{B}^{\ast} + \mathcal{B}) \psi_s  \right] \nonumber \\
i \partial_t \psi^{}_c &= \left[ - \frac{\nabla^2}{2m} + \omega_r (1+a_1 (\mathcal{Q})) +  4U_{01}|\psi^{}_0|^2 - \mathcal{E}_0 \right] \psi^{}_c
+ 2U_{01} \psi^{\ast}_c \psi^2_0 + U_{11} \psi^{}_c ( 3|\psi^{}_c|^2 + 2 |\psi^{}_s|^2 ) +U_{11} \psi^2_s \psi^{\ast}_c + 2 \eta \frac{w_0}{\sqrt{2}}(\mathcal{A}^{\ast} + \mathcal{A})\psi^{}_0\nonumber \\
i \partial_t \psi^{}_s &= \left[ - \frac{\nabla^2}{2m} + \omega_r (1+a_1 (\mathcal{Q})) +  4U_{01}|\psi^{}_0|^2 - \mathcal{E}_0 \right] \psi^{}_s
+ 2U_{01} \psi^{\ast}_s \psi^2_0 + U_{11} \psi^{}_s ( 3|\psi^{}_s|^2 + 2 |\psi^{}_c|^2 ) +U_{11} \psi^2_c \psi^{\ast}_s - 2 \eta \frac{w_0}{\sqrt{2}}(\mathcal{B}^{\ast} + \mathcal{B})\psi^{}_0\nonumber \\
i \partial_t \mathcal{A} &= -\Delta_C \left(1 - \frac{\nabla^2}{\zeta^2}\right) \mathcal{A} + 2\eta (\psi^{\ast}_{c} \psi^{}_0 + \psi^{\ast}_0 \psi^{}_c) \frac{w_0}{\sqrt{2}}  - i \kappa \mathcal{A}, \nonumber \\
i \partial_t \mathcal{B} &= -\Delta_C \left(1 - \frac{\nabla^2}{\zeta^2}\right) \mathcal{B} - 2\eta (\psi^{\ast}_{s} \psi^{}_0 + \psi^{\ast}_0 \psi^{}_s)\frac{w_0}{\sqrt{2}} \psi_0 - i \kappa \mathcal{B}.
\end{align}}
\hl{We note that $\mathcal{A}$ and $\mathcal{B}$ couple to $\psi_c$ and $\psi_s$ respectively. We also note there are two sources of anomalous coupling between $\psi_{c,s}$ and $\psi^\ast_{c,s}$; these come from the two sources of nonlinearity, i.e. the atomic interaction and indirectly via the coupling to the cavity field.} 

 \hl{These equations support various steady-state conditions.  There is always a steady state \hl{$\psi_{c}=\psi_{s}=\mathcal{A}=\mathcal{B}=0$}, corresponding to the normal state, below threshold.  In addition, above a critical pumping strength a DW polariton condensate state exists.
 We assume a uniform state, and without loss of generality we take $\psi_0$ to be real.  We may thus write:}
\hl{
\begin{equation}
\psi_{0S} = \sqrt{\rho_0},
~
\psi_{cS} = \sqrt{\rho_c}e^{i\phi_c}, 
~
\psi_{sS} = \sqrt{\rho_s} e^{i \phi_s},
~
\mathcal{A}_S = \frac{2\sqrt{2} \eta w_0 \sqrt{\rho_0 \rho_c}  \cos(\phi_c)}{\Delta_C + i \kappa}, 
~
\mathcal{B}_S = -\frac{2\sqrt{2} \eta w_0 \sqrt{\rho_0 \rho_s}  \cos(\phi_s)}{\Delta_C + i \kappa}.
\label{eq:ss_soln}
\end{equation}
}
\hl{
By considering the imaginary part of the steady-state equations for $\psi_{c,s}$, we find the following condition constraining the relative phases $\phi_{c,s}$:
\begin{align}
\left(\frac{1}{2} \chi_0-2 U_{01}\right) \rho_0 \sin(2 \phi_c) + U_{11} \rho_s \sin(2\phi_s - 2\phi_c) &= 0 \nonumber \\
\left(\frac{1}{2} \chi_0-2 U_{01}\right)\rho_0 \sin(2 \phi_s) - U_{11} \rho_c \sin(2\phi_s - 2\phi_c) &= 0.
\end{align}
where we have introduced the static cavity response
\be
\label{eq:chi0}
\chi_0 \equiv -\frac{8 \eta^2 w^2_0 \Delta_C}{\Delta^2_C + \kappa^2},
\ee
which appears via 
 $\sqrt{2} \eta w_0 (\mathcal{A}_S + \mathcal{A}^\ast_S) = - \chi_0 \sqrt{\rho_0 \rho_c} \cos(\phi_c)$, along with a similar expression for $\mathcal{B}$ fields.
As we will see below, the ratio of $\rho_{c,s}$ to $\rho_0$ is constrained by the real part of the self-consistency conditions.  Thus, we require the prefactors of $\rho_0$ and $\rho_{c,s}$ to vanish separately.  This leaves us with the allowed solutions
\be
\phi_{c,s} \in \{ 0, \pi/2, \pi, 3\pi/2\}.
\ee
Two of these correspond to real fields, and two to pure imaginary fields.  We may note that the pure imaginary case describes a density wave that does not couple to the cavity light, as $\cos(\phi_{c,s})=0$, and so these solutions cannot spontaneously arise under the conditions we consider. In the following, we thus restrict ourselves to the simplest real solution, $\phi_c=\phi_s=0$, which maximises the light-matter coupling strength. 
}

\hl{The condensate chemical potential,  $\mathcal{E}_0$---which serves as the energy reference for all atomic excited states---can be found by solving the steady-state equation for $\psi_0$.  For real fields $\psi_c,\psi_s$ we find
\be
\mathcal{E}_0 = \omega_r a_0(\mathcal{Q}) + 2U_{00} \rho_0 + (6U_{01} - \chi_0) (\rho_c + \rho_s).
\ee
}

\hl{
Using the above, along with density conservation $\rho_0+\rho_c+\rho_s=\rho$, we may then solve the steady-state condition for the density-wave amplitudes
$\rho_{c,s}$. With the purely real solution, we find the densities satisfy the equation
\be
\rho_c + \rho_s = \frac{\mu}{\tilde{U}},
\ee
where we have introduced $\mu$, which acts as the chemical potential of density-wave polaritons, and $\tilde{U}$ which describes their interactions. 
Note that this equation only constrains $\rho_c+\rho_s$, reflecting the $U(1)$ symmetry which is spontaneously broken by the density-wave solution.
The chemical potential
is given by:
\be
\mu = \chi_0 \rho - \omega_0 - (6U_{01} - 2U_{00})\rho.
\ee
Here, $\omega_0 = \omega_r[1 + a_1(\mathcal{Q})-a_0(\mathcal{Q})]$. The effective interaction strength is given by
\be
\tilde{U} =  2 \chi_0 + 3 U_{11} - 12U_{01}+2U_{00},
\ee
which contains contributions from both atomic repulsion and the nonlinearity from atom number conservation (hence the appearance of $\chi_0$).
}

Note that the pump is red-detuned from the cavity resonance and thus $\Delta_C < 0$, \hl{so $\chi_0>0$.}
The DW polariton condensate state exists only when $\mu>0$, which requires a sufficiently large pump strength $\eta$. \hl{Using the threshold pump strength $\eta_{\text{th}}$, we may also write $\mu = [(\eta/\eta_{\text{th}})^2 -1][\omega_0 + (6U_{01}-2U_{00}) \rho]$.
As noted above Eq.~\eqref{eq:defUeff}, $\omega_0 \gg U_{2D} \rho$, so the atomic interaction terms in this expression are a small correction to $\omega_0$.}
We recover the below-threshold state if we set $\mu=0$.

\subsection{Dispersion relation and speed of sound}

To derive the dispersion relation, we now expand around the stationary state by using the Bogoliubov--de Gennes parametrisation, considering a fluctuation with wavevector $\mbf{k}$ and (complex) frequency $\nu$.  Note that in our system, since we are considering \hl{$\psi_{c,s}$} as the envelope function of a density wave,
$ \mathbf{k}\cdot\hat{x} = k_\perp$, $\mathbf{k}\cdot\hat{y} = k_y$, and $\mathbf{k}\cdot\hat{z} = 0$:
\hl{
\begin{align}
\psi_0(\mbf{r},t) &= \sqrt{\rho_0} + h e^{-i(\mbf{k} \cdot \mbf{r} + \nu t)} + j^{*} e^{i(\mbf{k} \cdot \mbf{r} + \nu^\ast t)} \nonumber \\
\psi_c(\mbf{r},t) &= \sqrt{\rho_c} + u e^{-i(\mbf{k} \cdot \mbf{r} + \nu t)} + v^{*} e^{i(\mbf{k} \cdot \mbf{r} + \nu^\ast t)}, \nonumber \\
\psi_s(\mbf{r},t) &= \sqrt{\rho_s} + w e^{-i(\mbf{k} \cdot \mbf{r} + \nu t)} + q^{*} e^{i(\mbf{k} \cdot \mbf{r} + \nu^\ast t)}, \nonumber \\
\mathcal{A}(\mbf{r},t) &= \mathcal{A}_S +  c e^{-i(\mbf{k} \cdot \mbf{r} + \nu t)} + d^{*} e^{i(\mbf{k} \cdot \mbf{r} + \nu^\ast t)}, \nonumber \\
\mathcal{B}(\mbf{r},t) &= \mathcal{B}_S +  f e^{-i(\mbf{k} \cdot \mbf{r} + \nu t)} + g^{*} e^{i(\mbf{k} \cdot \mbf{r} + \nu^\ast t)}.
\end{align}
}
We seek to find how the allowed value(s) of $\nu$ depend on $\mbf{k}$. Inserting these into the equations of motion, keeping terms up to linear order in small fluctuations,  and  matching the Fourier components, the equations for the Bogoliubov--de Gennes coefficients are
\hl{
\begin{align}
\nu h &= \left[ \omega_r a_0(\mathcal{Q}) - \mathcal{E}_0 +  \frac{k^2}{2m} \right] h + 2U_{00} \rho_0 (2h+j) + U_{01} \left[ (\rho_c + \rho_s)(4h + 2j) + \sqrt{\rho_0 \rho_c}(8u+4v) + \sqrt{\rho_0 \rho_s}(8w+4q) \right] \nonumber \\
&\quad -\chi_0(\sqrt{\rho_c \rho_0} u + \sqrt{\rho_s \rho_0} w) - \frac{\chi(\nu,\mbf{k})}{2} \left[ \sqrt{\rho_c \rho_0} (u+v) + \sqrt{\rho_s \rho_0}(w+q) + (\rho_c + \rho_s)(h+j) \right] \nonumber \\
-\nu j &= \left[ \omega_r a_0(\mathcal{Q}) - \mathcal{E}_0 +  \frac{k^2}{2m} \right] j + 2U_{00} \rho_0 (2j+h) + U_{01} \left[ (\rho_c + \rho_s)(4j + 2h) + \sqrt{\rho_0 \rho_c}(8v+4u) + \sqrt{\rho_0 \rho_s}(8q+4w) \right] \nonumber \\
&\quad -\chi_0(\sqrt{\rho_c \rho_0} v + \sqrt{\rho_s \rho_0} q) - \frac{\chi(\nu,\mbf{k})}{2} \left[ \sqrt{\rho_c \rho_0} (u+v) + \sqrt{\rho_s \rho_0}(w+q) + (\rho_c + \rho_s)(h+j) \right] \nonumber \\
\nu u &= \left[\omega_r(1+a_1(\mathcal{Q})) -\mathcal{E}_0+ 4 U_{01} \rho_0 + \frac{k^2}{2m} \right] u + 4 U_{01} \sqrt{\rho_0 \rho_c}(h+j) + 2 U_{01}( \rho_0 v + 2 \sqrt{\rho_0 \rho_c} h) \nonumber \\
&\quad +U_{11} [(3 \rho_c + \rho_s)(2u+v) + 2 \sqrt{\rho_s \rho_c}(2w + q)]  - \chi_0 \sqrt{\rho_0 \rho_c} h - \frac{\chi(\nu,\mbf{k})}{2} \left[ \rho_0(u+v) + \sqrt{\rho_0 \rho_c} (h+j) \right], \nonumber \\
-\nu v &= \left[ \omega_r(1+a_1(\mathcal{Q})) -\mathcal{E}_0+ 4 U_{01} \rho_0 +  \frac{k^2}{2m} \right] v + 4 U_{01} \sqrt{\rho_0 \rho_c}(h+j) + 2 U_{01}( \rho_0 u + 2 \sqrt{\rho_0 \rho_c} j) \nonumber \\ &\quad +U_{11} [(3 \rho_c + \rho_s)(2v+u) + 2 \sqrt{\rho_s \rho_c}(2q + w)] - \chi_0 \sqrt{\rho_0 \rho_c} j - \frac{\chi(\nu,\mbf{k})}{2} \left[ \rho_0(u+v) + \sqrt{\rho_0 \rho_c} (h+j) \right]\nonumber \\
\nu w &= \left[\omega_r(1+a_1(\mathcal{Q})) -\mathcal{E}_0+ 4 U_{01} \rho_0 +  \frac{k^2}{2m} \right] w + 4 U_{01} \sqrt{\rho_0 \rho_s}(h+j) + 2 U_{01}( \rho_0 q + 2 \sqrt{\rho_0 \rho_s} h) \nonumber \\
&\quad +U_{11} [(3 \rho_s + \rho_c)(2w+q) + 2 \sqrt{\rho_s \rho_c}(2u + v)] - \chi_0 \sqrt{\rho_0 \rho_s} h - \frac{\chi(\nu,\mbf{k})}{2} \left[ \rho_0(w+q) + \sqrt{\rho_0 \rho_s} (h+j) \right] \nonumber \\
-\nu q &= \left[ \omega_r(1+a_1(\mathcal{Q})) -\mathcal{E}_0+ 4 U_{01} \rho_0 +  \frac{k^2}{2m} \right] q + 4 U_{01} \sqrt{\rho_0 \rho_s}(h+j) + 2 U_{01}( \rho_0 w + 2 \sqrt{\rho_0 \rho_s} j) \nonumber \\ &\quad +U_{11} [(3 \rho_s + \rho_c)(2q+w) + 2 \sqrt{\rho_s \rho_c}(2v + u)] - \chi_0 \sqrt{\rho_0 \rho_s} j - \frac{\chi(\nu,\mbf{k})}{2} \left[ \rho_0(w+q) + \sqrt{\rho_0 \rho_s} (h+j) \right].
\end{align}
Here,  we have introduced the dynamic cavity response
\begin{equation}
\chi(\nu,\mbf{k}) = \frac{8 \eta^2 w^2_0 \Delta_k}{\Delta^2_k - (\nu + i\kappa)^2}, \qquad
\Delta_k \equiv -\Delta_C(1 + k^2/\zeta^2).
\end{equation}
in addition to the static cavity response $\chi_0$ defined in Eq.~\eqref{eq:chi0}.
}
\hl{
This cavity response comes from eliminating $c,d,f,g$ from the atomic equations by expressing them in terms of the atomic fluctuations, using
\begin{align}
\nu c &= \left[\Delta_k - i \kappa\right] c + \frac{2\eta w_0 }{\sqrt{2}}\left[\sqrt{\rho_0}(u+v) + \sqrt{\rho_c}(h+j)\right], \quad
-\nu d = \left[\Delta_k + i \kappa\right] d + \frac{2\eta w_0}{\sqrt{2}}\left[\sqrt{\rho_0}(u+v) + \sqrt{\rho_c}(h+j)\right],
\nonumber \\
\nu f &= \left[\Delta_k - i \kappa\right] f - \frac{2\eta w_0}{\sqrt{2}}\left[\sqrt{\rho_0}(w+q) + \sqrt{\rho_s}(h+j)\right], \quad
-\nu g = \left[\Delta_k + i \kappa\right] g - \frac{2\eta w_0}{\sqrt{2}}\left[\sqrt{\rho_0}(w+q) + \sqrt{\rho_s}(h+j)\right].
\end{align}
}

\hl{
The equations for atomic fluctuations can be re-arranged into the matrix form
\be
\begin{bmatrix}
A_0(\nu) - \nu &  B_0(\nu) & C_c(\nu) & D_c(\nu) & C_s(\nu) & D_s(\nu) \\
B_0(\nu) & A_0(\nu) + \nu & D_c(\nu) & C_c(\nu) & D_s(\nu) & C_s(\nu)  \\
C_c(\nu) & D_c(\nu) & A_c(\nu) - \nu &  B_c(\nu) & 2E & E \\
D_c(\nu) & C_c(\nu) & B_c(\nu ) & A_c(\nu) + \nu & E & 2E \\
C_s(\nu) & D_s(\nu) & 2E & E & A_s(\nu) - \nu &  B_s(\nu) \\
D_s(\nu) & C_s(\nu) & E & 2E & B_s(\nu) & A_s(\nu) + \nu
\end{bmatrix}
\begin{bmatrix}
h \\
j \\
u \\
v \\
w \\
q \\
\end{bmatrix}
=0,
\ee
where for $i = c,s$, (and after using the definition of $\mathcal{E}_0$), we find:
\begin{align}
A_0(\nu) &\equiv 2U_{00}\rho_0 + \left[ \chi_0 - 2U_{01} - \frac{\chi(\nu,\mbf{k})}{2} \right](\rho_c + \rho_s) + \frac{k^2}{2m}, \nonumber \\
B_0(\nu) &\equiv 2U_{00}\rho_0 + \left[ 2U_{01} - \frac{\chi(\nu,\mbf{k})}{2} \right](\rho_c + \rho_s), \nonumber \\
A_i(\nu) &\equiv \omega_0 +  \left[ 4U_{01}-2U_{00} - \frac{\chi(\nu,\mbf{k})}{2}\right] \rho_0  + (2 U_{11}-6U_{01} + \chi_0)(\rho_c + \rho_s) + 4 U_{11} \rho_i +\frac{k^2}{2m}, \nonumber \\
B_i(\nu) &\equiv \left[2 U_{01} - \frac{\chi(\nu,\mbf{k})}{2} \right] \rho_0  + U_{11}(\rho_c +  \rho_s) + 2U_{11}\rho_i \nonumber \\
C_i(\nu) &\equiv \sqrt{\rho_0 \rho_i} \left[ 8U_{01} - \chi_0  - \frac{\chi(\nu,\mbf{k})}{2} \right], \nonumber \\
D_i(\nu) &\equiv \sqrt{\rho_0 \rho_i} \left[ 4U_{01} - \frac{\chi(\nu,\mbf{k})}{2} \right] \nonumber \\
E &\equiv 2 U_{11} \sqrt{\rho_s \rho_c}.
\end{align}
Below threshold, where $\rho_{c,s}=0$, this structure simplifies into three $2\times 2$ blocks, as $C_i=D_i=E=0$ in this case.  The dispersion of the density-wave polariton modes is given by the middle or lower-right blocks, which both have the same spectrum.  Above threshold the off-diagonal elements are non-zero, making the problem more complicated.
}

\hl{
The structure of these equations can however be considerably simplified by a change of parameterisation, so that we consider fluctuations of the density-wave envelope which are in-phase or out-of-phase with the steady state.  The in-phase mode corresponds to amplitude fluctuations and so is expected to be gapped, while the out-of-phase mode corresponds to displacement of the atomic density wave, and so should be gapless.
To make this change of basis, we first define the variable $\varphi$ by parameterising the steady-state solution as $\rho_c = (\rho - \rho_0) \cos^2(\varphi),~\rho_s = (\rho - \rho_0) \sin^2(\varphi)$. Our basis change then corresponds to a rotation matrix:
\be
\begin{bmatrix}
h \\
j \\
u \\
v \\
w \\
q \\
\end{bmatrix} = 
\begin{bmatrix}
1 & 0 & 0 & 0 & 0 & 0\\
0 & 1 & 0 & 0 & 0 & 0\\
0 & 0 & \cos(\varphi) & 0 & -\sin(\varphi) & 0\\
0 & 0 & 0 & \cos(\varphi) & 0 & -\sin(\varphi) \\
0 & 0 & \sin(\varphi) & 0 & \cos(\varphi) & 0 \\
0 & 0 & 0 & \sin(\varphi) & 0 & \cos(\varphi) \\
\end{bmatrix}
\begin{bmatrix}
h^{\prime} \\
j^{\prime} \\
u^{\prime} \\
v^{\prime} \\
w^{\prime} \\
q^{\prime} \\
\end{bmatrix}.
\ee
This rotation puts
the matrix equation  into block-diagonal form
\be
\begin{bmatrix}
A_0(\nu)-\nu & B_0(\nu) & C(\nu) & D(\nu) & 0 & 0 \\
B_0(\nu) & A_0(\nu) + \nu & D(\nu) & C(\nu) & 0 & 0 \\
C(\nu) & D(\nu) & A_+(\nu) - \nu &  B_+(\nu) & 0 & 0 \\
D(\nu) & C(\nu) & B_+(\nu ) & A_+(\nu) + \nu & 0 & 0 \\
0 & 0 & 0 & 0 & A_-(\nu) - \nu &  B_-(\nu) \\
0 & 0 & 0 & 0 & B_-(\nu) & A_-(\nu) + \nu
\end{bmatrix}
\begin{bmatrix}
h^{\prime} \\
j^{\prime} \\
u^{\prime} \\
v^{\prime} \\
w^{\prime} \\
q^{\prime} \\
\end{bmatrix} = 0,
\label{eq:r_mat_eq}
\ee
with new coefficients:
\begin{align}
A_{\pm}(\nu) &= \omega_0 +\frac{k^2}{2m} + \left[4U_{01} - 2U_{00} -\frac{\chi(\nu,\mbf{k})}{2} \right] \rho_0  + (4U_{11}-6U_{01}+\chi_0)(\rho_c + \rho_s) \pm 2U_{11}(\rho_c + \rho_s) \nonumber \\
B_{\pm}(\nu) &= \left[ 2U_{01} - \frac{\chi(\nu,\mbf{k})}{2} \right] \rho_0 + 2 U_{11} (\rho_c + \rho_s) \pm U_{11} (\rho_c + \rho_s) \nonumber \\
C(\nu) &= \sqrt{\rho_0(\rho_c + \rho_s)} \left[8U_{01} - \chi_0 -\frac{\chi(\nu,\mbf{k})}{2} \right] \rho_0  \nonumber \\
D(\nu) &= \sqrt{\rho_0(\rho_c + \rho_s)}\left[ 4U_{01} -\frac{\chi(\nu,\mbf{k})}{2} \right] .
\end{align}
This structure naturally reveals a simplified  lower-right $2 \times 2$ block that describes the gapless phase mode, while the remaining $4\times 4$ block describes gapped amplitude oscillations.  Note that this reparameterisation is also valid (albeit unnecssary) below threshold.}

\subsubsection{Below threshold}

\hl{
Below threshold, there is no atom population in the scattered states and therefore $\rho_{c,s} = 0, ~\rho_0 = \rho$. The excitation spectrum can be found from solutions of the equation:
\be
B_-(\nu)^2 = [A_-(\nu) - \nu][A_-(\nu) + \nu].
\label{det_eq}
\ee
}
Focusing on the experimentally relevant limit $\nu, \kappa \ll |\Delta_C|$,  we obtain the simple expression
\hl{
\be
\nu(\mbf{k}) = \sqrt{\left[\omega_0 +\frac{k^2}{2m}+ (6U_{01} - 2U_{00})\rho  + \frac{8 \eta^2 N}{\Delta_C(1 + k^2/\zeta^2)}\right]\left[ \omega_0 + \frac{k^2}{2m}  + 2(U_{01} - U_{00})\rho \right]},
\ee
}%
\hl{where $N = \rho w^2_0/2$ is the total number of the atoms}. This expression is the dispersion of DW polaritons: the normal-mode frequencies result from mixing the atomic density wave dispersion $k^2/2m$ with the photon-mediated interaction.  As observed in the main text, this expression exhibits a relatively flat dispersion (controlled by the atomic mass) when $\eta=0$.  Because $\Delta_C<0$, increasing $\eta$ both softens the mode---reduces $\nu(\mbf{k}=0)$---and leads to a steeper dispersion, due to the $\mbf{k}$-dependence of the cavity-mediated term.   When \hl{$8 \eta^2 N = - \Delta_C [\omega_0 + (6U_{01} - 2U_{00})\rho]$}, one has $\nu(\mbf{k}=0)=0$; this corresponds to the point at which the mode becomes entirely soft, and DW polariton condensation occurs. (Note, this condition \hl{is written assuming} $\kappa \ll |\Delta_C|$, as introduced above).

\subsubsection{Above threshold}

\begin{figure*}
    \includegraphics[width = 0.45\textwidth]{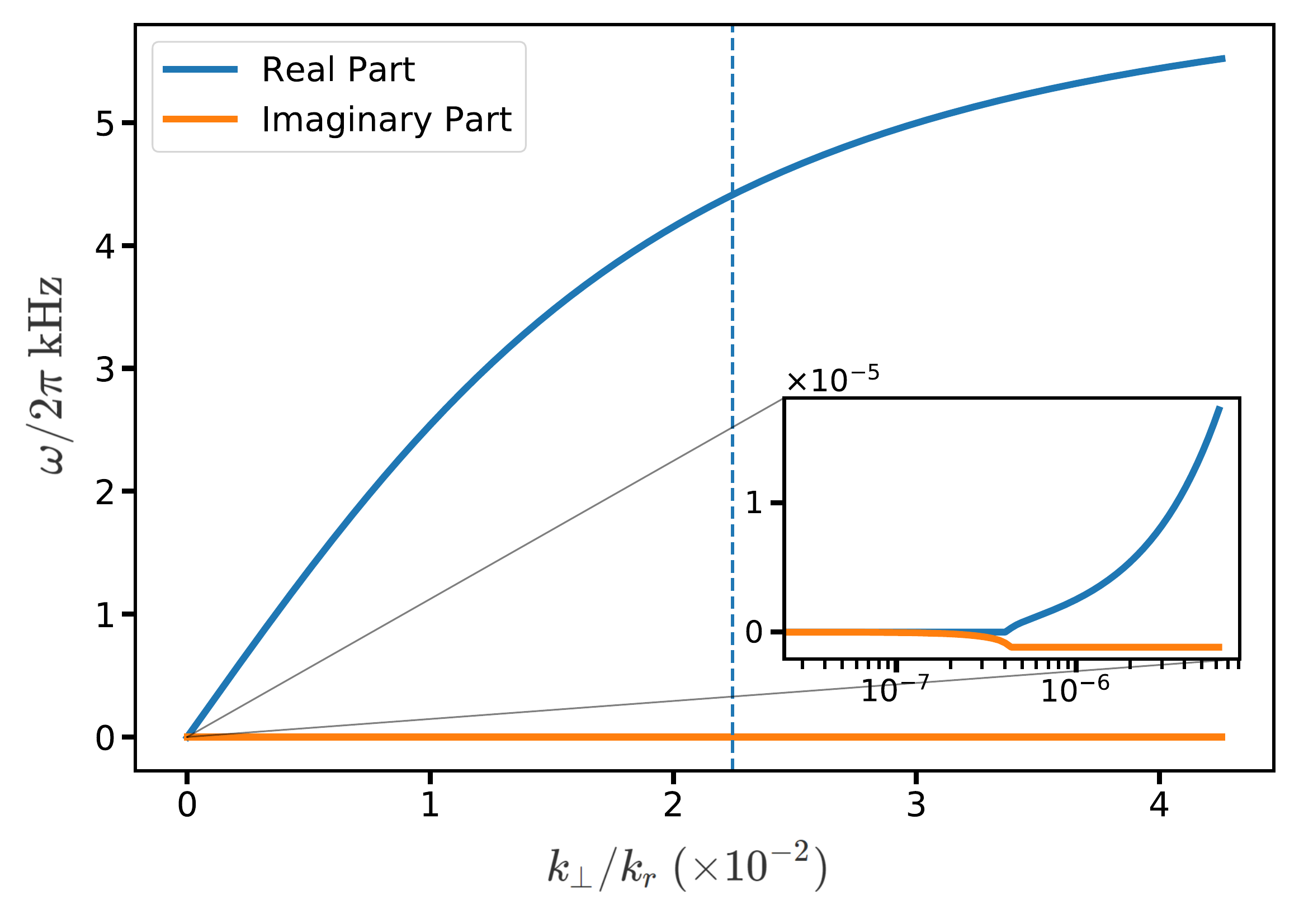} 
    \caption{Numerical solution to the dispersion above threshold. The cavity dispersion flattens toward a Debye frequency of ${\sim}$5~kHz at large $k_\perp$. Vertical dashed line indicates $k_{\perp} = \zeta$. Inset shows the spectrum at small momenta.}
    \label{fig:full_theory}
\end{figure*}

Above threshold, \hl{we use the same determinantal equation Eq.~\eqref{det_eq}, along with the steady-state solution $\rho_c + \rho_s = \mu/\tilde{U}$, to find the normal mode frequencies}. To obtain a simple expression, we expand \hl{the term $\chi(\nu,\mbf{k})$ in $A_-(\nu)$ and $B_-(\nu)$} up to linear order in $\nu/|\Delta_k|$ and $\kappa/|\Delta_k|$, which is the relevant limit for the long-wavelength, low-energy regime we explore:
\hl{
\be
\chi(\nu,\mbf{k}) = \frac{8 \eta^2 w^2_0}{\Delta_k }\left[1 + \frac{2 i \nu \kappa}{\Delta^2_k}\right] + O\left(\left|\frac{\nu}{\Delta_k}\right|^2,\left|\frac{\kappa}{\Delta_k}\right|^2\right).
\ee
}
The determinantal equation for $\nu$ in Eq.~\eqref{det_eq} can then be shown to have the solution:
\hl{
\be
\label{eq:DWpoldispersion}
\nu(\mbf{k}) = i \frac{\Gamma_k}{2} \pm \sqrt{\mathcal{M}_k\mathcal{N}_k - \frac{\Gamma_k^2}{4}},
\ee
where 
\begin{align}
\Gamma_k &= -\frac{16 w^2_0 \rho_0\mathcal{M}_k \eta^2 \kappa}{\Delta^3_k} \nonumber \\
\mathcal{M}_k &= \omega_0 + \frac{k^2}{2m} + 2(U_{01} - U_{00}) \rho + (U_{11} - 8 U_{01} + 2 U_{00} + \chi_0) \frac{\mu}{\tilde{U}} \nonumber \\
\mathcal{N}_k &= \frac{k^2}{2m} - \frac{8 \eta^2 w^2_0 \rho_0}{\Delta_C} \left[ 1 - \frac{1}{1+k^2/\zeta^2}\right].
\end{align}
}
The $\Gamma_k$ expression is the excitation gap due to cavity loss, \hl{and one may see that $\mathcal{N}_k$ vanishes at ${\mbf k}\to 0$, leading to a gapless spectrum.}

Because $\kappa \ll |\Delta_C|$, the dissipative term $\Gamma_k$ has a small effect except at very small $k$.  Outside this small $k$ regime, the real part of the dispersion can be expanded up to linear order to yield the dispersion relation of the acoustic \hl{phonon} mode:
\hl{
\be
\Re[\nu (\mbf{k})] 
\simeq \sqrt{\mathcal{M}_0 \mathcal{N}_k}
= v_s |\mbf{k}|,
\ee
where the speed of sound takes the form
\be
v_s = \sqrt{\mathcal{M}_0 \left( \frac{1}{2m} + \frac{E_I}{\zeta^2} \right) },
\ee
}
in terms of the cavity-mediated interaction energy scale
$E_I = -{8N \eta^2}/{\Delta_C}$. One may note that $E_I/\zeta^2 \gg 1/2m$ \hl{in the presence of a deep pump lattice}, so the sound velocity is principally determined by the cavity-mediated interaction strength.  \hl{For the factor $\mathcal{M}_0$, we may note that, as previously discussed, $\omega_0  \gg U_{2D} \rho$. From the definition of threshold, we note further that above threshold one may also approximate $\chi_0  \gg U_{2D}$.  This means we may approximate $\mathcal{M}_0\simeq \omega_0 + \mu/2 = (\omega_0/2)[(\eta/\eta_{\text{th}})^2 + 1]$.
When the pump is not too strong, one may note that $\omega_0\simeq 2 \omega_r$, giving the result quoted in the main text.
} 
For the above-threshold data presented in the main text, $v_s \approx 16$~cm/s.

Using the above expression, we can also describe the diffusive
regime that occurs at very small momenta.  We use that
\hl{$\mathcal{M}_k \mathcal{N}_k\simeq v_s^2 k^2$} at small momentum to find that as $\mbf{k} \to 0$  the gapless branch of $\nu(\mbf{k})$ has the purely imaginary diffusive spectrum $\nu(\mbf{k}) = i 2 v_s^2 k^2 / \Gamma_k$, as also found for microcavity polariton condensates~\cite{Szymanska2006nqc,Carusotto2013qfo}. This crosses over to the linear sound dispersion when
$v_s k = \Gamma_k/2$. Figure~\ref{fig:full_theory} plots the solution~\eqref{eq:DWpoldispersion} to the above-threshold determinant equation using experimental parameters. The imaginary part contribution $\propto\Gamma_k$ is indeed negligible.

As discussed above, one should note that $k_\perp$ is measured from the $(\pm k_r,\pm k_r)$ points.  In the above equations, this can be seen from the fact that $\psi_1(k)$ is the envelope function multiplying an atomic density wave with wavevector $k_r$. We also note one particular higher-order effect that is not included in the above treatment.  The missing effect arises from the fact that photon momentum must be conserved, $k^2 = k_\parallel^2 + k_\perp^2 = 4\pi^2/\lambda^2$. Consequently, phonons with nonzero $k_\perp$ propagate in supermodes with a concomitantly reduced $k_\parallel$. Nevertheless, this effect is negligible for the momenta considered in the current experiment ($k_{\perp}/k_r \sim 10^{-2}$). For the largest momentum measured, the fractional change in the longitudinal momentum $k_{\parallel}$ is less than $10^{-4}$.

We also note that there should be a phonon dispersion in $\hat{y}$ as well as that which we have shown in $\hat{x}$.  We do not attempt to measure the dispersion in $\hat{y}$ because we choose to make a BEC that is thin in this direction so as to maximise its length in $\hat{x}$.  That is, the BEC is too thin in $\hat{y}$ to support a full wavelength of the shortest-wavelength phonons we are able to stimulate at present.

\section{Derivation of the form of the light profile emitted from the cavity}

We now derive the relationship of the intracavity field to that emitted by the cavity.  Moreover, we describe how a phonon excitation at a particular momentum $k_\perp$ appears in holographic images of the cavity field. We first show how each cavity field couples to momentum excitations. To do so, we consider the equation of motions in Eq.~\eqref{EOM} for $\alpha_\mu \equiv \langle \hat{a}_{\mu} \rangle$ and $\beta_\mu \equiv \langle \hat{b}_{\mu} \rangle$
\begin{align}
        i \partial_t \alpha_{\mu} &= -(\Delta_{\mu} + i\kappa) \alpha_{\mu} - \eta \sqrt{2}  \int d \mbf{r} \sum_{\mu}\Xi_{\mu}(\mbf{r})\left\{\hat{\psi}^{\dagger}_0(\mbf{r}) \left[\hat{\psi}^{}_{c}(\mbf{r})  +i\hat{\psi}^{}_{s}(\mbf{r})\right] e^{-i n_\mu \theta_0} + \hat{\psi}^{}_0(\mbf{r}) \left[\hat{\psi}^{\dagger}_{c}(\mbf{r})  +i\hat{\psi}^{\dagger}_{s}(\mbf{r})\right] e^{-i n_\mu \theta_0} + \text{H.c.} \right\} \nonumber \\
    i \partial_t \beta_{\mu} &= -(\Delta_{\mu} + i \kappa) \beta_{\mu} - \eta \sqrt{2}  \int d \mbf{r} \sum_{\mu}\Xi_{\mu}(\mbf{r})\left\{i\hat{\psi}^{\dagger}_0(\mbf{r}) \left[\hat{\psi}^{}_{c}(\mbf{r})  +i\hat{\psi}^{}_{s}(\mbf{r})\right] e^{-i n_\mu \theta_0} + i\hat{\psi}^{}_0(\mbf{r}) \left[\hat{\psi}^{\dagger}_{c}(\mbf{r})  +i\hat{\psi}^{\dagger}_{s}(\mbf{r})\right] e^{-i n_\mu \theta_0} + \text{H.c.} \right\}
\end{align}

\hl{We note that because, as discussed above, the relevant steady-state has real $\psi_c$ and $\psi_s$, we can therefore  consider $\psi_c, \psi_s$ to be the real part and the imaginary part of a complex field $\psi_1 = \psi_c + i \psi_s$. Resolving $\psi_1$ into Fourier components
\be
\psi_1(\mbf{r}) = \int d\mbf{k} \psi_{\mbf{k}} e^{-i \mbf{k} \cdot \mbf{r}},
\ee
we can consider the light profile due to the atomic population in a particular momentum mode $\psi_{\mbf{k}}$ by rewriting the equations of motions in terms of $\psi_{\mbf{k}}$
\begin{align}
    i \partial_t \alpha_{\mu} &= -(\Delta_{\mu} + i\kappa) \alpha_{\mu} - \eta \sqrt{2} \int d \mbf{k} \int d\mbf{r} \Xi_{\mu}(\mbf{r}) [ \psi_{0} (\mbf{r})\psi^{*}_{\mbf{k}} e^{i (\mbf{k} \cdot \mbf{r} + n_\mu \theta_0)}+ \text{H.c.}],  \\
    i \partial_t \beta_{\mu} &= -(\Delta_{\mu} + i \kappa) \beta_{\mu} - \eta \sqrt{2} \int d \mbf{k} \int d \mbf{r} \Xi_{\mu}(\mbf{r}) [\psi_{0} (\mbf{r}) \psi^{*}_{\mbf{k}} i e^{i (\mbf{k} \cdot \mbf{r} + n_\mu \theta_0)}+ \text{H.c.}].  
\end{align}
}

Focusing on the first degenerate resonance with even modes, and after setting the time derivative to zero, the light profile is
\begin{align}
\tilde{\alpha} (\mbf{r},z) &= \sum_{\mu} \alpha_{\mu} \Xi_{\mu} (\mbf{r}) \cos(k_r z - \theta_0 - \Theta - n_{\mu} \theta_0) \nonumber \\
&= -\sum_{\mu}\hl{\sqrt{2}}\eta  \int d\mbf{r}^\prime \frac{\Xi_{\mu}(\mbf{r}) \Xi_{\mu}(\mbf{r}^\prime)}{\Delta_{\mu} + i \kappa} \mathcal{S}^{+}_{\mu}\psi_0 (\mbf{r}^\prime) \int d\mbf{k} [\psi^{*}_{\mbf{k}}  e^{i (\mbf{k} \cdot \mbf{r}^\prime + n_\mu \theta_0)}+ \text{H.c.}]\cos(k_r z - \theta_0 - \Theta - n_{\mu} \theta_0),
\end{align}
where the tilde signifies that this includes the profile along $\hat{z}$.
Taking the forward-travelling part of the standing-wave intracavity field, we find the form of the cavity emission out of one of its cavity mirrors (for one of the two cavity fields):
%\begin{widetext}
\begin{align}\label{firstforward}
\tilde{\alpha}^{+} (\mbf{r},z) &\propto e^{i(k_r z - \theta_0 - \Theta)} \sum_{\mu}\int d\mbf{r}^\prime \frac{\Xi_{\mu}(\mbf{r}) \Xi_{\mu}(\mbf{r}^\prime)}{\Delta_{\mu} + i \kappa} \mathcal{S}^{+}_{\mu}\psi_0 (\mbf{r}^\prime) \int d\mbf{k} [\psi^{*}_{\mbf{k}} e^{i \mbf{k} \cdot \mbf{r}^\prime} + \psi_{\mbf{k}} e^{-i(\mbf{k} \cdot \mbf{r}^\prime + 2 n_{\mu} \theta_0)}] \nonumber \\
&\propto e^{i(k_r z - \theta_0 - \Theta)} \int d\mbf{k} \int d\mbf{r}^\prime \psi_0 (\mbf{r}^\prime) \left[\mathcal{G}^{+} (\mbf{r},\mbf{r}^\prime,0) \psi^{*}_{\mbf{k}} e^{i \mbf{k} \cdot \mbf{r}^\prime} + \mathcal{G}^{+} (\mbf{r},\mbf{r}^\prime,-2i\theta_0) \psi_{\mbf{k} }e^{-i \mbf{k} \cdot \mbf{r}^\prime}  \right],
\end{align}
%\end{widetext}
where $\mathcal{G}^{+}(\mbf{r},\mbf{r}^\prime,\varphi)$ is a modified Green's function \hl{defined now with the dissipation term, given by $i\kappa$, explicit}~\cite{Vaidya2017tpa}
\begin{align}
\mathcal{G}^{+}(\mbf{r},\mbf{r}^\prime,\varphi) &= \mathcal{G}(\mbf{r},\mbf{r}^\prime,\varphi) + \mathcal{G}(\mbf{r},-\mbf{r}^\prime,\varphi), \nonumber \\
\mathcal{G}(\mbf{r},\mbf{r}^\prime,\varphi) &= \displaystyle\sum_{\mu} \frac{\Xi_{\mu}(\mbf{r})\Xi_{\mu}(\mbf{r}^\prime) e^{-n_\mu \varphi}}{1 + \tilde{\epsilon} n_\mu + i \tilde{\kappa}}.
\end{align}
The calculation for $\beta_{\mu}$ proceeds in the same way, with the only difference being the aforementioned additional $\pi/2$ phase shift:
%\begin{widetext}
\begin{align}
\tilde{\beta} (\mbf{r},z) &= \sum_{\mu} \beta_{\mu} \Xi_{\mu} (\mbf{r}) \sin(k_r z - \theta_0 - \Theta - n_{\mu} \theta_0) \nonumber \\
&= -\sum_{\mu}\eta \int d\mbf{r}^\prime \frac{\Xi_{\mu}(\mbf{r}) \Xi_{\mu}(\mbf{r}^\prime)}{\Delta_{\mu} + i \kappa} \psi_0 (\mbf{r}^\prime) \int d\mbf{k} [i\psi^{*}_{\mbf{k}}  e^{i (\mbf{k} \cdot \mbf{r}^\prime + n_\mu \theta_0)}+ \text{H.c.}]\sin(k_r z - \theta_0 - \Theta - n_{\mu} \theta_0), \\
\tilde{\beta}^{+} (\mbf{r},z) &\propto e^{i(k_r z - \theta_0 - \Theta)} \sum_{\mu}\int d\mbf{r}^\prime \frac{\Xi_{\mu}(\mbf{r}) \Xi_{\mu}(\mbf{r}^\prime)}{\Delta_{\mu} + i \kappa} \psi_0 (\mbf{r}^\prime) \int d\mbf{k} [\psi^{*}_{\mbf{k}} e^{i \mbf{k} \cdot \mbf{r}^\prime} - \psi_{\mbf{k}} e^{-i(\mbf{k} \cdot \mbf{r}^\prime + 2 n_{\mu} \theta_0)}] \nonumber \\  \label{secondforward}
&\propto e^{i(k_r z - \theta_0 - \Theta)} \int d\mbf{k} \int d\mbf{r}^\prime \psi_0 (\mbf{r}^\prime) \left[\mathcal{G}^{+} (\mbf{r},\mbf{r}^\prime,0) \psi^{*}_{\mbf{k}} e^{i \mbf{k} \cdot \mbf{r}^\prime} - \mathcal{G}^{+} (\mbf{r},\mbf{r}^\prime,-2i\theta_0) \psi_{\mbf{k} }e^{-i \mbf{k} \cdot \mbf{r}^\prime}  \right]. 
\end{align}
%\end{widetext}
After measuring the cavity field emission from both frequencies, the total field can be reconstructed digitally by summing Eqs.~\ref{firstforward} and~\ref{secondforward}:
\be
\Phi(\mbf{r}) \propto \int d\mbf{k} \int d\mbf{r}^\prime \psi_0 (\mbf{r}^\prime) \mathcal{G}^{+} (\mbf{r},\mbf{r}^\prime,0) \psi^{*}_{\mbf{k}} e^{i \mbf{k} \cdot \mbf{r}^\prime}. 
\label{camera_field}
\ee
Therefore, the digitally summed image contains the contribution from only the local part $\mathcal{G} (\mbf{r},\mbf{r}^\prime,0)$ of the field because the nonlocal contribution from $\mathcal{G} (\mbf{r},\mbf{r}^\prime,-2i\theta_0)$ cancels, as shown in Fig.~\ref{fig2} in the main text.

In the limit of an ideal confocal cavity in which  $\mathcal{G} (\mbf{r},\mbf{r}^\prime,0)$ becomes a $\delta$-function (i.e., perfect mode degeneracy), Eq.~\eqref{camera_field} shows that a single momentum component $\psi_{\mbf{k}}$ will result in a phase winding $e^{i \mbf{p} \cdot \mbf{r}}$ on the reconstructed cavity emission. Taking the Fourier transform of the complex electric field $\Phi(\mbf{r})$ then reveals the momentum mode that has been stimulated through Bragg spectroscopy. Furthermore, this also shows that a particular momentum can be excited by stimulating the local part of the field and probing on either one of the degenerate resonances. 

Though we do not employ this here, we note that the nonlocal part of the cavity emission---i.e., the part that could be found by digitally subtracting the images at the two cavity frequencies---also provides information about the driven momentum. This is because, for a confocal cavity with a BEC confined to the cavity midpoint $z=0$, this emission is the Fourier transform of the object image. Consider the simple case of an ideal confocal cavity with atoms located at the midplane of the cavity where $\theta_0 = \pi/4$.   The nonlocal part of the emitted field is
\begin{align}
\Phi_{\mrm{nonlocal}}(\mbf{r}) &\propto \int d\mbf{k} \int d\mbf{r}^\prime \psi_0 (\mbf{r}^\prime) \cos\left[ \frac{2 \mbf{r} \cdot \mbf{r}^\prime}{w^2_0}\right] \psi^{*}_{\mbf{k}} e^{i \mbf{k} \cdot \mbf{r}^\prime} \nonumber \\
&= \frac{1}{2} \int d\mbf{k} \int d\mbf{r}^\prime \psi_0 (\mbf{r}^\prime) \psi^{*}_{\mbf{k}} \left\{ \mathrm{exp} \left[ i \left(   \mbf{k} + \frac{2 \mbf{r}}{w^2_0} \right) \cdot \mbf{r}^\prime \right] + \mathrm{exp} \left[ i \left(   \mbf{k} - \frac{2 \mbf{r}}{w^2_0} \right) \cdot \mbf{r}^\prime \right]  \right\},
\label{nonlocal_field}
\end{align}
which contains direct information about the excited momentum modes $\psi_{\mbf{k}}$.

\section{Measurement of temporal dynamics}
\label{sec:time_dynamics}
\begin{figure*}[t]
  \includegraphics[width=\textwidth]{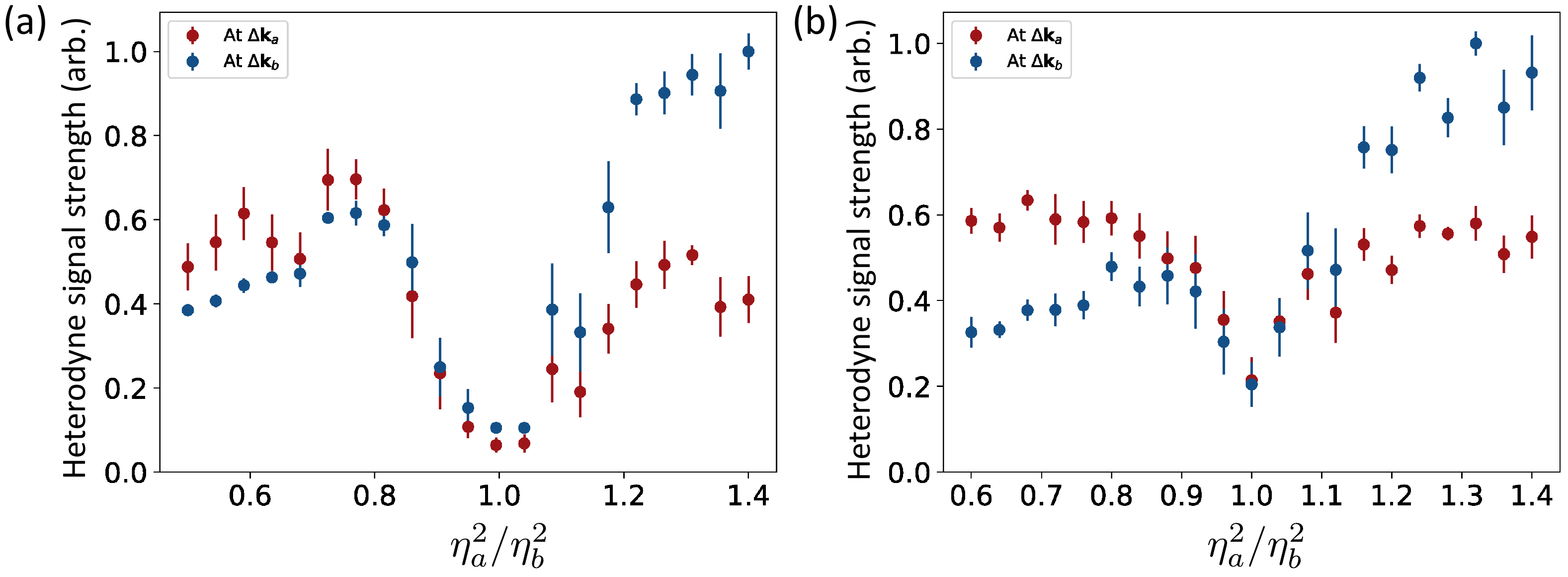}
  \caption{\hl{Spatial heterodyne signal as a function of the ratio of strengths of the two transverse pumping beams. Plotted for (a) a 5-ms  camera integration time and (b) a 2-ms integration time.}  }
  \label{SuppFig_balance}
\end{figure*}

\hl{While the intensity time-trace shown in Fig.~\ref{fig2}a shows no obvious above-threshold dynamics, it is  not sensitive to the phase of the intracavity field. In contrast, the holographic reconstruction does capture some dynamical information:  The time-evolution of $\Delta\phi_i(\mbf{r})+\delta_i$  during the EMCCD integration time leads to the reduction of fringe contrast, as measured by $\chi_i$. (See Sec.~\ref{holosec} for notation definitions.) Physically, this happens when the above-threshold DW starts to slide due to the $U(1)$ degree-of-freedom along the cavity axis.} 

\hl{We can qualitatively investigate this by recording the total weight of the Fourier components in a region around $\Delta\mbf{k}_a$ and $\Delta\mbf{k}_b$, respectively, versus the imbalance in pump power $\eta^2_a/\eta^2_b$.
When the pump beams are balanced, the full $U(1)$ longitudinal phase symmetry is achieved and the DW is free to slide on timescales shorter than the integration time. This serves to  wash-out the interference fringes.
When the powers are not balanced, the cancellation of the non-local interaction is imperfect, lifting the  translational symmetry along the cavity axis. This lack of translationally symmetry then leads to the pinning of the longitudinal phase of the DW. This leads to a significant increase in fringe contrast.}

\hl{Figure~\ref{SuppFig_balance} shows this for two different pump ramp rates. For a 5~ms integration time, shown in panel (a), one sees a notable dip in the Fourier weights when $\eta_a\simeq\eta_b$.
We repeated this measurement for a faster pump ramp of 2~ms and  EMCCD integration time of 2~ms; see panel (b). The faster rate gives the DW dynamics less time to dephase the interference fringes, and we observe a dip both narrower and shallower in contrast. }

\section{Linear phase gradient in Figure 2 images}

\hl{As noted in the main text, a rainbow-like baseline linear phase gradient, corresponding to a shear of the emergent lattice, is seen in the images of the supermode DW. 
In this section we briefly discuss the apparent origin of this phase gradient. First, though, we emphasize that this phase gradient coexists with the lattice showing an overall $U(1)$ symmetry under displacements---the results in Fig.~\ref{fig2}f were taken at $\eta^2/\eta^2_{\mrm{th}}=10$, for which the gradient is appreciable.
Since this gradient does not pin the $U(1)$ phase, it cannot affect the dispersion of low-energy phonons either: the properties of such phonons are symmetry-protected. Therefore this phase gradient does not affect our analysis above, except potentially by weakly renormalizing the speed of sound.}

\begin{figure*}[t]
  \includegraphics[width=0.45\textwidth]{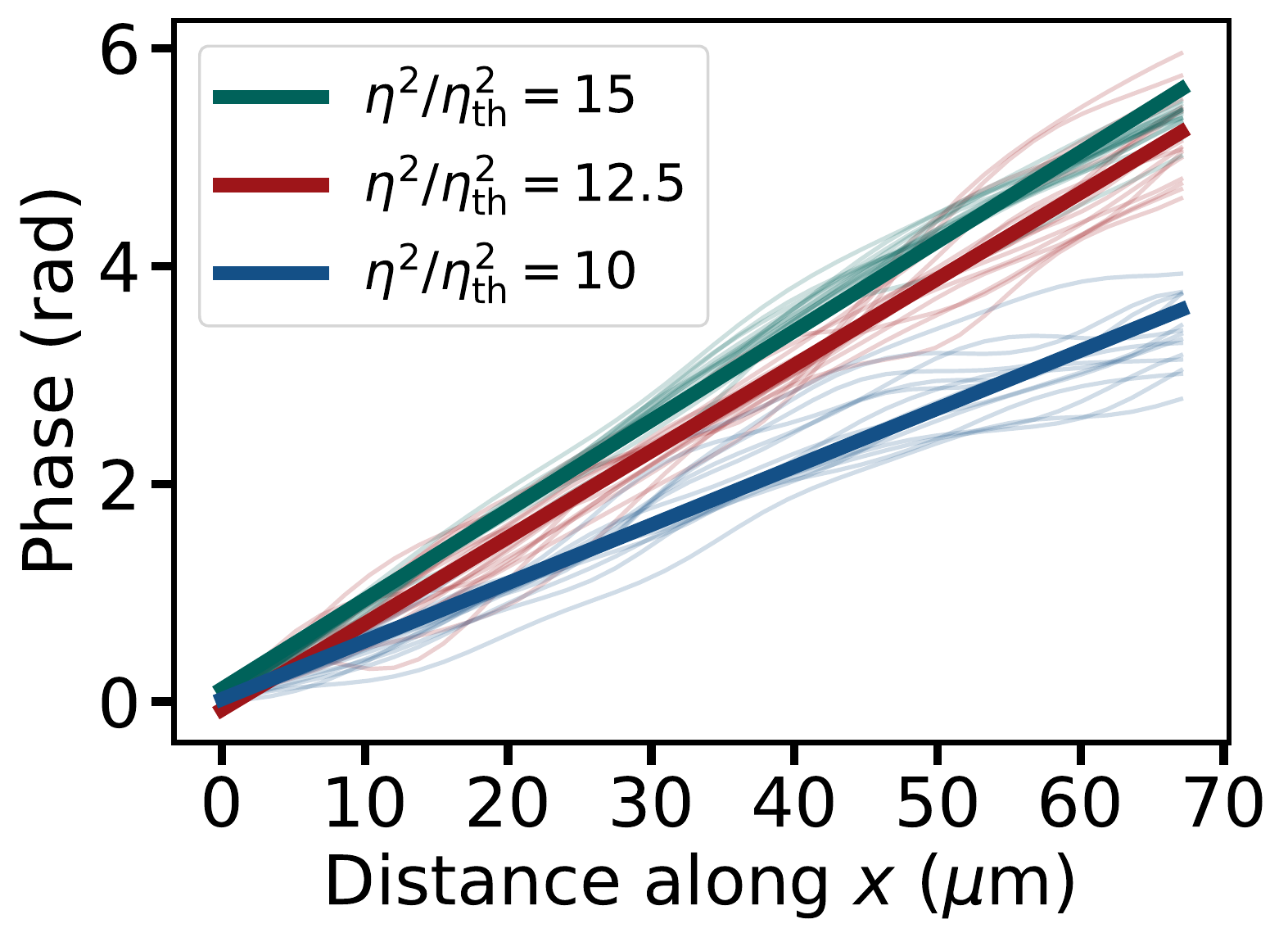}
  \caption{Linear background gradients at different final ramp powers measured by line traces in cavity emission images, such as that in Fig.~\ref{fig2}b,c. Semitransparent lines are measured data and solid lines are the best linear fit for each set.}
  \label{SuppFig_gradient}
\end{figure*}

\hl{Nevertheless, the consistent appearance of this gradient implies there must be some systematic cause.  In this section we show evidence that the origin of this gradient is a nonlinear effect, which occurs at large pumping strength.  
As evidence, in Fig.~\ref{SuppFig_gradient} we show the phase gradient found by crossing the threshold and ramping pump power to $\eta^2/\eta^2_{\mrm{th}} = 10$, 12.5, and 15. One sees that the phase gradient becomes larger at greater pump power.
While the gradient could also include an artefact of a small-angle (${\sim}0.45^\circ$) misalignment between the BEC and $\hat{x}$, Fig.~\ref{SuppFig_gradient} indicates that the dominant contribution is pump strength dependent. }

\hl{
Strong pumping ($\eta^2/\eta^2_\text{th}\gtrsim 10$) is needed to obtain the illustrative, high-signal-to-noise images and data shown in Fig.~\ref{fig2} and Fig.~\ref{SuppFig_gradient}.  As such, it is not possible to fully image how the phase gradient evolves as one approaches the threshold.   In contrast, strong pumping is not needed for the phonon dispersion measurements reported in main text.  In that case, we employ far weaker ($\eta^2/\eta^2_\text{th} \leq 1.25$), near-threshold pump strengths.  The results of Fig.~\ref{SuppFig_gradient} suggest any linear gradient would be much weaker in such a regime. }

\section{Imaging long-wavelength phonon}\label{imagingsec}

\begin{figure*}
 \includegraphics[width=\textwidth]{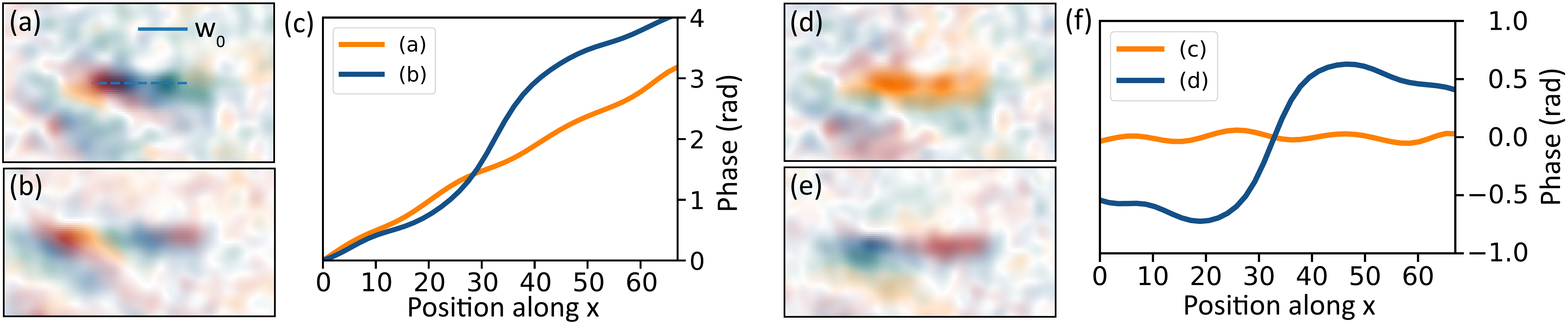}
  \caption{\hl{Image of an imprinted phonon $\propto e^{i k_\perp x}$.  Raw data is shown in panels (a-c). We have removed the background linear phase gradient in panels (d-f).  (a) Raw image of the  DW polariton without a longitudinal field present. Dashed line indicates where the line traces in panels (c) and (f) are typically taken. (b) Raw image of the  DW polariton stimulated by a longitudinal field with $k_\perp =0.08$~rad$/\mu$m $=1{\times}10^{-2}k_r$.  (c) Line traces from the data in panels (a,b), where the orange (blue) line is without (with) the longitudinal pump. Image (d) is without, while (e) is with the longitudinal field. Both have been processed to remove the background linear phase gradient. (f) The nearly flat orange line shows no phonon imprinted.  The blue line shows a phase trace with a phase modulation arising from the longitudinal field.  We observe roughly a full-wavelength of the expected DW modulation before the onset of BEC finite-size effects.  We note that there is an arbitrary phase offset in these measurements.  We choose to set this phase reference to zero at $x=0$ in panel (c), but we choose to set it to zero at the middle of the image in panel (f). }}
  \label{PhononImage}
\end{figure*}

\hl{To demonstrate the imaging of an imprinted phonon, we cross the supermode DW polariton condensation threshold in the presence of a longitudinal field of fixed wavevector.  We then ramp the pump strength to be well above the threshold to obtain the emission images in Fig.~\ref{PhononImage}.  Like a ferromagnetic transition in the presence of an inhomogeneous longitudinal magnetic field, this breaks the underlying symmetry `by hand,' stimulating the emergent lattice into one particular phonon mode.  With no longitudinal field, the cavity emission reveals a flat phase profile (after the background linear phase gradient is removed).   By contrast, when we imprint a longitudinal field with a particular $k_\perp$, we observe a phase modulation. }

We also note that these images are taken by first exciting the phonon modes near the pump power $\eta^2 \approx \eta^2_\text{th}$ and then ramping up one pump to be much stronger than the other for enhanced contrast for the image associated with the LO from that pump. The beam power is rapidly ramped up to $\eta^2/\eta^2_\text{th}\approx10$ in a duration of $500~\mu$s, which is much faster than the phonon dynamics.

\section{Coupling of a longitudinal probe field into the confocal cavity}

The Hamiltonian for a longitudinally driven multimode cavity is given by
\be
H= -\sum_{\mu}\left[\Delta_{\mu} \hat{a}^\dagger_{\mu} \hat{a}^{}_{\mu} + f_{\mu} (\hat{a}_{\mu}  + \hat{a}^{\dagger}_{\mu})\right],
\ee
where \be
f_{\mu} = \int d\mbf{r}^\prime f(\mbf{r}^\prime) \Phi_{\mu}(\mbf{r},^\prime z_0)  
\ee
is the spatial overlap between the longitudinal pump $f(\mbf{r})$ and a cavity mode $\Phi_{\mu}(\mbf{r},z)$ near a particular plane $z=z_0$ is
\be
 \Phi_{\mu}(\mbf{r},z_0)=
 \Xi_{\mu} (\mbf{r}) \cos[k_r z_0 - \theta_\mu(z_0)].
\ee
Ignoring cavity loss, the classical equation of motion of the expectation value $\alpha_\mu = \langle \hat{a}_{\mu} \rangle$ is
\be
i \partial_t \alpha_{\mu} = \Delta_{\mu} \alpha_{\mu} + f_{\mu}.
\ee
In steady state, the total transverse light field is therefore given by
\be
\alpha(\mbf{r}) = \sum_{\mu} \alpha_{\mu} \Phi_{\mu}(\mbf{r},z_0) = \sum_{\mu} \frac{f_\mu \Phi_{\mu}(\mbf{r},z_0)}{\Delta_{\mu}}.
\ee
Focusing on the transverse profile, the cavity field is
\be
\alpha(\mbf{r}) = \int d\mbf{r}^\prime f(\mbf{r}^\prime)\sum_{\mu} \frac{\Xi_{\mu}(\mbf{r})\Xi_{\mu}(\mbf{r}^\prime) \cos^2\left[k_r z_0 -\theta_{\mu}(z_0)\right]}{\Delta_{\mu}} \mathcal{S}^{+}_{\mu} \equiv \int d\mbf{r}^\prime f(\mbf{r}^\prime) \mathcal{T} (\mbf{r}, \mbf{r}^\prime),
\ee
where $\theta_{\mu}$ is the longitudinal phase of mode $\mu$ at the transverse plane $z = z_0$ we are considering. The factor $\mathcal{S}^{+}_{\mu}$ restricts the summation to modes with even transverse spatial symmetry for the case of a degenerate resonance in a confocal cavity. Thus, the transfer function $\mathcal{T}$ for an longitudinally input beam profile can be evaluated the same way as the Green's function for the cavity-mediated interaction. 

In an ideal confocal cavity, at the cavity midplane, the transfer function contains three parts~\cite{Vaidya2017tpa}:
\be
\mathcal{T}(\mbf{r},\mbf{r}^\prime) \propto \delta\Big(\frac{\mbf{r} - \mbf{r}^\prime}{w_0/\sqrt{2}}\Big) + \delta\Big(\frac{\mbf{r} + \mbf{r}^\prime}{w_0/\sqrt{2}}\Big) + \frac{1}{\pi} \cos \Big( \frac{2\mbf{r} \cdot \mbf{r}^\prime}{w_0^2}\Big).
\ee
For a longitudinal probe profile localised around the atoms, the resulting cavity field will contain the probe field, its mirror image, and a nonlocal contribution. In a realistic cavity where the transverse modes are not fully degenerate (such as ours), the highest spatial-frequency components of the input probe field will be suppressed.

\section{Measuring Spontaneous Symmetry Breaking of $U(1)$ symmetry}
\hl{We now discuss how we measured the shot-to-shot phase fluctuations presented in Fig.~\ref{fig2}(f). In principle, the breaking of  $U(1)$ symmetry  can be directly measured from the phase difference between the LO beam and the cavity emission.   This phase difference can be extracted from the image in a spatial heterodyne measurement, which would manifest as an overall shot-to-shot phase shift in the interference fringes in the entire image. However, the shot-to-shot relative phase between the LO and the cavity emission suffers from technical drift due to fluctuations in optical path lengths. Nevertheless, in our previous work of Ref.~\cite{Guo2019eab}, we  showed that in a confocal cavity, the spatial phase difference $\Delta \phi$ between the local and nonlocal part of the cavity field is directly related to the phase of the atomic DW along the cavity axis $\phi_A$ via the relation $\phi_A = -2 \Delta \phi$. With this approach of measuring relative spatial phase in cavity emission, the technical phase drift is reduced to an overall phase shift on both the local part and nonlocal part of the field that drops out of their difference. Computing the quantity $\Delta \phi$ in the spatial heterodyne image then cancels this overall phase drift, and the atomic density-wave phase $\phi_A$ can be measured from shot-to-shot.} 

\hl{As discussed  in Sec.~\ref{sec:time_dynamics} regarding time dynamics, when the power of two pump beams is balanced, the phase of the atomic density wave is allowed to freely slide along the cavity axis, which results in a significant reduction in the spatial heterodyne signal strength for integration time longer than around 2 ms. As such, to achieve a reasonable level of signal-to-noise ratio for measuring the phase of the cavity field for both the local and non-local components, we rapidly ramp up the pump power to $\eta^2/\eta^2_{\mathrm{th}} = 10$ in 750 $\mu$s, faster than the typical timescale of the phonon dynamics. Additionally, we employ a $\sim$35~$\mu$m-long gas for higher atomic density for further enhancing the cavity emission field amplitude. As shown in Fig.~\ref{fig2}c in the main text, the phase of the local part of the field is taken from the peak amplitude of the electric field localized around the atoms, while the phase of the nonlocal part is computed from an average of a patch of the electric field off to the side of the atoms.}

\section{Comparison to other systems}

In a crystallisation transition, the mutual interactions of the particles in a gas or liquid conspire to break a continuous translational $U(1)$ symmetry. While the periodicity of the resultant lattice is set by properties of the particles themselves, the phase of the lattice freely emerges.  Goldstone's theorem implies that after solidification, the lattice may, in principle, slide at no energy cost since no particular phase had been preferred~\cite{Altland2006cmf}. While this zero-momentum ($k=0$) mode contributes nothing to the thermodynamic properties of the solid, long-wavelength (small $k$) phonon modes that do contribute also arise from the broken symmetry. These connect with the $k=0$ mode to form a continuous, gapless spectrum of excitations called a Goldstone mode.

While the amplitude of the lattice is emergent in a single-mode, single-pumped cavity, its phase and periodicity are geometrically fixed by the single cavity mode into which the photons scatter---the lattice remains inelastic. Adding a second, frequency-degenerate cavity mode creates a standing-wave potential for the atoms with an emergent phase, thereby breaking the $U(1)$ symmetry.  As the atoms move, so does the standing wave, realising the $k=0$ point of a Goldstone mode. Such a scheme was experimentally realized using two crossed single-mode Fabry-P\'{e}rot cavities tuned to the same frequency~\cite{Leonard2017sfi}, \hl{and more recently in a ring cavity~\cite{Schuster2020spo}.} Pumped atoms scatter photons into a superposition of the two modes. Their coherent sum yields a $U(1)$ phase degree of freedom and realises a lattice with the same $k=0$ mode mentioned above.  With the atoms Bose-condensed,  a simple supersolid is created wherein superfluidity and periodic structure arising from the broken translational symmetry coexist. Absent, however, are phonons. This is because the infinite-range, photon-mediated interactions of each single-mode cavity \hl{(and the ring cavity)} yield an effectively 0D system with no $k>0$ modes~\cite{Leonard2017mam,Lang2017cea}---the Goldstone dispersion is missing. A similar $k=0$ Goldstone mode of a dipolar supersolid has been observed~\cite{Guo2019tlg}.

\section{Data availability statement}

The datasets generated during the current study are available in the Harvard Dataverse Repository, https://doi.org/10.7910/DVN/LGT5O6.

\section{Author information statement}

The authors declare no competing interests.

\section{Additional information}
Supplementary Information is available for this paper. Correspondence and requests for materials should be addressed to Benjamin Lev at benlev@stanford.edu. Reprints and permissions information is available at www.nature.com/reprints.

\end{document}